\documentclass[journal]{IEEEtran}

\usepackage{cite}
\usepackage{amsmath,amssymb,amsfonts}
\usepackage{graphicx}
\usepackage{textcomp}
\usepackage{url}
\usepackage{booktabs}
\usepackage{multirow}
\usepackage{array}
\usepackage[below]{placeins}

\newcommand{\R}{\mathbb{R}}

\newsavebox{\TableBody}
\newcounter{algorithm}
\newcommand{\TableNotes}[2]{\parbox{#1}{\footnotesize\raggedright\textit{Notes:} #2}}

\title{Structured Speaker Conditioning for Diarization-Guided Target-Speaker ASR}

\author{Ruoyu~Wang, Shutong~Niu, Jun~Du, Weixin~Zhu, and Yannan~Wang
\thanks{Ruoyu Wang, Shutong Niu, and Jun Du are with the University of Science and
Technology of China, Hefei, China (e-mail: wangruoyu@mail.ustc.edu.cn,
jundu@ustc.edu.cn).}
\thanks{Weixin Zhu and Yannan Wang are with Ethereal Audio Lab, China.}}

\begin{document}
\maketitle

\begin{abstract}
We propose structured speaker conditioning (SSC) for diarization-guided
target-speaker recognition. SSC combines local activity structure with
session-level acoustic identity to construct identity-consistent target
conditions that control the acoustic encoder. Target-centric relational
encoding (TCRE) retains per-speaker activity histories and factorized second-
and third-order relations, capturing overlap and turn changes with
aggregation cost linear in the number of tracks. Reference-based speaker
conditioning (RSC) supplies session-derived identities that disambiguate
local activity patterns within TCRE. Conditional low-rank modulation (CLRM)
converts the joint condition into channel-wise affine modulation and a
low-rank cross-channel residual with a condition-dependent core.
On Qwen3-ASR-0.6B with oracle diarization, SSC uses 1.46M conditioning
parameters and achieves a 6.23\% mean relative reduction in concatenated
minimum-permutation error rate (cpER) over four-state diagonal conditioning
across four meeting test sets. SSC also transfers across audio foundation
models. With predicted diarization and target-error group sequence policy
optimization, the FireRedAudio text system achieves mean relative error
reductions of 17.9\% and 17.3\% over DiCoW and MOSS-TD on the same four
sets, and also outperforms both on three unseen domains.
Scenario analyses identify
gains under challenging overlap conditions, while fixed-checkpoint
interventions support the contributions of session identity and the
condition-dependent core. A timestamp variant jointly produces sentence
boundaries and text. With predicted diarization, macro-average cpER across
the four sets is 15.60\% for text and 15.96\% with timestamps.
\end{abstract}


\section{Introduction}
\label{sec:introduction}
\IEEEPARstart{M}{eeting} transcription should recover what each participant
said while maintaining consistent speaker attribution across turn changes
and overlapping speech. Target-speaker automatic speech recognition (ASR)
addresses this problem by generating a transcript of a specified speaker's
speech from the mixture. When a long meeting is processed in windows, the
condition should keep referring to the same session speaker as the local
combination of active speakers changes. \emph{How can a target condition remain identity-consistent under
changing local activity and ambiguous speaker cues while effectively
controlling the acoustic encoder?}

Meeting transcription systems combine diarization, separation, and recognition
in modular pipelines~\cite{yoshioka2019advances,vonneumann2024meeting}, or
jointly predict speaker identities and transcripts~\cite{kanda2020joint}.
Recent speech-language models also generate timestamps, as in
MOSS-TD~\cite{mosi2026moss} and TagSpeech~\cite{huo2026tagspeech}.
Diarization-guided target-speaker ASR directly supplies
speaker-activity tracks to a recognition model. DiCoW
\cite{polok2024target,polok2025dicow} demonstrates this approach by adapting
Whisper~\cite{radford2023whisper} with frame-level
diarization-dependent transformations (FDDT), and
Dixtral~\cite{groundingspoken2026} carries this conditioning into a spoken
language model with a frozen language decoder.
For each target, the activity tracks are converted into four states:
silence, target-only, non-target-only, and overlap (STNO). Their probabilities
control affine transformations of the acoustic features. This provides a
basis for studying how diarization should be represented
and how the condition should act on a pre-trained encoder.

The information retained in the condition determines which speaker
configurations the encoder can distinguish. Four-state conditioning can map
different interferer counts and hand-over histories to the same target
condition. Encoding individual track histories before aggregation preserves
these distinctions, including how the target and other speakers overlap
and alternate over time.

Acoustic identity complements this temporal structure. Two speakers with
identical activity throughout the available context receive the same
activity-only condition, while speech elsewhere in the session can provide
acoustic cues for distinguishing their voices. SE-DiCoW~\cite{polok2026sedicow} introduces self-enrollment
through encoder cross-attention, and SC-SOT~\cite{hirano2025scsot} conditions
the decoder on speaker embeddings and activity. Combining identity with
track histories within a frame-level condition lets both sources determine
the target--interferer relations used for recognition.

The modulation operator then translates these relations into acoustic
adaptation. Diagonal FDDT scales and shifts channels, while self-speaker
adaptation~\cite{wang2025selfspeaker} generates speaker-specific encoder
kernels from activity supervision. A low-rank cross-channel operator offers
a compact way to vary channel combinations through a small generated core
and shared read/write projections. Jointly designing the condition and this
operator lets target identity and local speaker interactions control how
acoustic features are combined.

We address this question with \emph{structured speaker conditioning}
(SSC). Target-centric relational encoding (TCRE) retains per-speaker activity
histories and their overlap and turn-taking relations. Reference-based
speaker conditioning (RSC) incorporates speaker identity information derived
from session audio and diarization into these track representations, so
TCRE can distinguish targets with similar local activity. Conditional
low-rank modulation (CLRM) uses the joint condition to control channel-wise
adaptation and cross-channel feature combinations in the acoustic encoder.

\begin{itemize}
  \item \textbf{TCRE} constructs permutation-equivariant target conditions
  using a shared causal temporal
  encoder and factorized second- and third-order relations. The representation
  captures activity history, interferer counts, and interactions with distinct
  interferer pairs. We relate its design to an analysis of STNO's activity compression.
  \item \textbf{RSC} associates session-derived acoustic identities with
  diarization tracks and augments TCRE features with this identity evidence.
  It supports target recognition when target speech is fully overlapped,
  target activity is missed, or speaker tracks are swapped.
  \item \textbf{CLRM} combines feature-wise
  linear modulation (FiLM) \cite{perez2018film} with a rank-8 cross-channel
  residual whose core varies with the target condition. The
  operator extends diagonal modulation with condition-dependent
  channel mixing. Operator comparisons and frozen-core interventions link
  this design to lower error.
\end{itemize}

SSC is trained using supervised fine-tuning (SFT). The FireRedAudio-based
systems are subsequently optimized using group sequence policy optimization
(GSPO)~\cite{zheng2025gspo}, with rewards based on target-speaker
transcription error.

We evaluate SSC on Qwen3-ASR and FireRedAudio, using four
in-domain meeting test sets (AliMeeting, AISHELL-4, AMI-SDM, and NOTSOFAR-1)
and three out-of-domain test sets. Recognition performance is evaluated
using concatenated minimum-permutation error rate (cpER), calculated at the
character level for Chinese and the word level for English. On
Qwen3-ASR-0.6B with oracle diarization, complete SSC uses
1.46M conditioning parameters and achieves a mean relative cpER reduction
of 6.23\% over STNO conditioning with the diagonal FDDT operator
(STNO-FDDT).
With predicted diarization, FireRedAudio + SSC + GSPO achieves
mean relative cpER reductions of 17.9\% and 17.3\% over DiCoW and
MOSS-TD, respectively, across the four in-domain sets. Scenario-specific
analyses show larger gains in windows with at least three concurrent speakers
and on fully overlapped target speech, while condition interventions show that session identity
contributes most under target misses and track swaps.
A timestamp-output variant jointly generates sentence boundaries and
transcripts for recognition and timing evaluation.

\section{Related Work}
\label{sec:related_work}
\subsection{Meeting Transcription and Target-Speaker Recognition}
Modular meeting transcription combines separation, recognition, and speaker
attribution. Yoshioka et al.\ \cite{yoshioka2019advances} integrate continuous
speech separation into an online separate--recognize--diarize system.
Permutation-invariant training \cite{yu2017pit} addresses the output-order
ambiguity of speaker-independent separation, while TF-GridNet
\cite{wang2023tfgridnet} models local and global time--frequency structure.
Building on TF-GridNet separation, von Neumann et al.\
\cite{vonneumann2024meeting} combine ASR with transcription-supported
diarization. These pipelines expose separated speech and speaker assignment
as intermediate representations.

Joint recognition approaches learn to generate multiple speakers' transcripts
from the mixture. Serialized output training (SOT) \cite{kanda2020sot}
concatenates their transcriptions in a single decoder sequence.
Speaker-attributed extensions~\cite{kanda2020joint} add a speaker inventory
and jointly model speaker counting, transcription, and identification.
Token-level SOT \cite{kanda2022tsot} interleaves recognition tokens and marks switches between
virtual output channels for streaming recognition. Segment-based SOT
\cite{subramanian2025segsot} further addresses practical online and offline
decoding, and LLM-based SOT \cite{shi2024llmsot} combines a pretrained
speech encoder with a language-model decoder. Recent speech-language
models extend structured generation to speaker labels and timestamps.
MOSS-TD \cite{mosi2026moss} and VibeVoice-ASR \cite{peng2026vibevoice}
jointly generate these outputs for long recordings, while DiaScriber
\cite{mu2026diascriber} adapts a speech language model for joint diarization
and transcription. TagSpeech~\cite{huo2026tagspeech} uses interleaved time
anchors to synchronize semantic and speaker streams for temporal grounding.
Target-speaker ASR supplies an explicit condition
specifying whose speech to decode. SSC adopts this formulation and uses
session-level diarization to associate window-level predictions with meeting
speakers.

\subsection{Activity and Acoustic Identity as Speaker Conditions}
Speaker activity provides temporal information for target recognition.
End-to-end neural diarization directly estimates overlapping speaker
activities \cite{fujita2019eend}; encoder--decoder attractors extend this
formulation to an unknown number of speakers \cite{eendeda2020}.
TS-VAD \cite{tsvad2020} instead conditions activity estimation on speaker
representations. These diarization approaches supply activity estimates;
target-speaker recognition uses such estimates to guide transcription.
Whisper-based target-speaker ASR~\cite{polok2024target,polok2025dicow}
conditions the acoustic encoder through STNO-weighted feature
transformations. In the diagonal configuration, these transformations map
activity states to channel-wise gains and offsets. Dixtral~\cite{groundingspoken2026} couples
the DiCoW encoder with Voxtral, adapting the acoustic encoder and FDDT
while freezing the language decoder and modality adapter.

\begin{figure}[!t]
\centering
\includegraphics[width=\columnwidth]{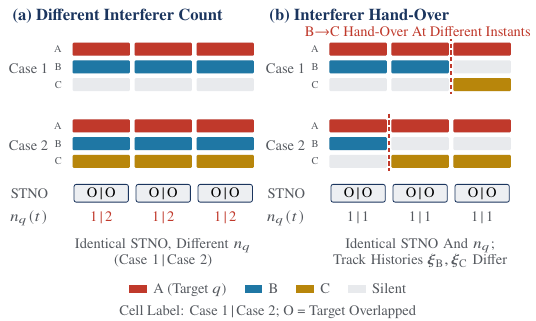}
\caption{STNO activity compression. Different interferer counts (a)
or hand-over histories (b) can yield the same four-state target
condition despite distinct non-target track configurations.}
\label{fig:stno_collision}
\end{figure}

Fig.~\ref{fig:stno_collision} illustrates the information compressed
by this four-state representation. With binary tracks, one or two
simultaneous interferers can yield identical STNO conditions. Different
hand-over times between non-target speakers can also leave the STNO
sequence unchanged despite distinct per-track histories. These cases motivate
encoding track histories before symmetric aggregation and preserving
relations between the target and individual interferers.

Self-speaker adaptation~\cite{wang2025selfspeaker} generates speaker-specific
kernels from speaker-activity supervision and injects them into selected
encoder layers. At the decoder, SC-SOT~\cite{hirano2025scsot} combines
speaker embeddings from a jointly trained diarization model with activity
information. DM-ASR~\cite{li2026dmasr} converts speaker labels and segment
boundaries into prompts for speaker- and time-conditioned dialogue turns.
SSC combines explicit target--interferer relations with conditional
low-rank modulation in the encoder.

Reference speech supplies acoustic identity complementary to activity.
Speaker embedding models such as x-vectors \cite{snyder2018xvectors} and
ECAPA-TDNN \cite{desplanques2020ecapa} summarize utterances as identity
representations.
SpeakerBeam \cite{zmolikova2019speakerbeam} learns a target representation
from an adaptation utterance and uses it to condition speaker extraction.
VoiceFilter \cite{wang2019voicefilter} combines a speaker embedding with
spectrogram masking for targeted separation. For recognition, Kanda et al.\
\cite{kanda2019simultaneous} alternate speaker-embedding estimation with
target-speaker ASR, obtaining identity evidence from the dialogue itself.
SQ-Whisper \cite{sqwhisper2024} uses trainable queries to
derive enrollment-guided speaker prompts from the mixture, and supplies the
prompts to Whisper's encoder and decoder. SE-DiCoW
\cite{polok2026sedicow} selects an enrollment segment from the conversation
using diarization and introduces its encoded feature sequence through
cross-attention at each encoder layer, while retaining the STNO path.

Deep Sets~\cite{zaheer2017deepsets} uses symmetric aggregation over unordered
inputs, while Relation Networks~\cite{santoro2017relations} aggregate
learned pairwise relations. In SSC, RSC identities augment temporal track
features before TCRE constructs target-centric second- and third-order
relations. The target identity also enters the final
condition directly, jointly controlling diagonal modulation and low-rank
cross-channel mixing.

\section{Method}
\label{sec:method}

\begin{figure*}[!t]
\centering
\includegraphics[width=\textwidth]{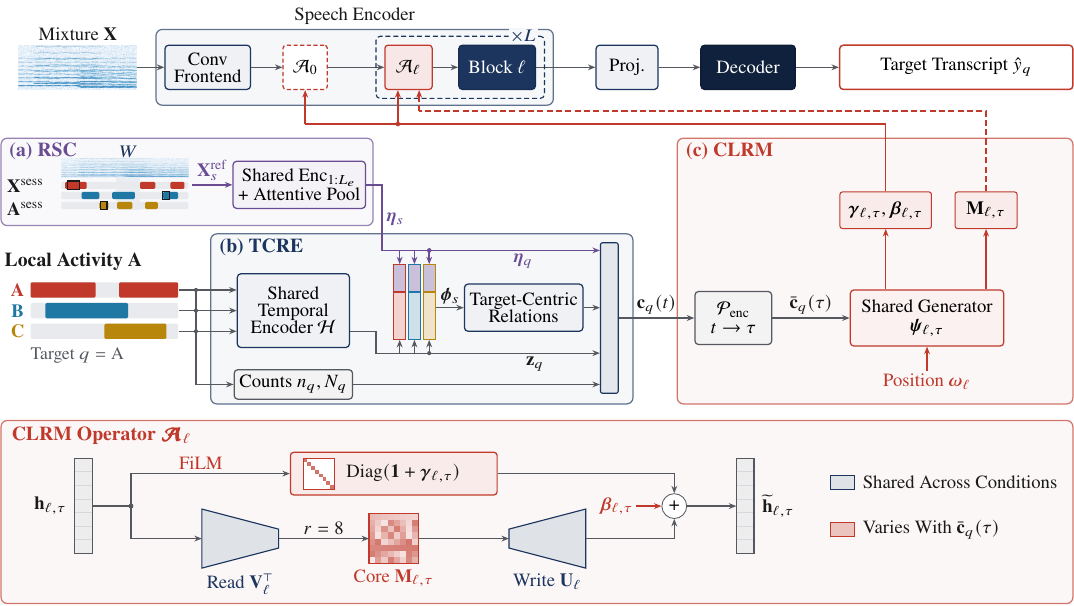}
\caption{SSC architecture. (a) RSC derives speaker identities from session
audio and diarization. (b) TCRE combines these identities with local track
histories into $\mathbf c_q(t)$. (c) CLRM aligns the condition to encoder
tokens and generates modulation parameters. The enlarged operator below
combines diagonal FiLM with a low-rank read--mix--write branch on the same
acoustic feature. Read/write bases are shared across conditions; the core
varies with the condition. The pre-stack position $\ell=0$ uses FiLM only.}
\label{fig:ssc_framework}
\end{figure*}

\subsection{Task Definition and Conditioning Pipeline}
Let $\mathbf{X}\in\R^{T\times F}$ denote the acoustic features of a meeting
window $W$, and let $\mathbf{A}=[a_s(t)]\in[0,1]^{S\times T}$ contain its
speaker-activity tracks on the same frame grid. Each $a_s(t)$ is the fraction
of a frame covered by diarization intervals for speaker $s$. Track indices
are associated with session speakers, and $q\in\{1,\ldots,S\}$ identifies
the requested speaker. Given session audio $\mathbf{X}^{\mathrm{sess}}$ and diarization
$\mathbf{A}^{\mathrm{sess}}$, RSC derives a set of acoustic identities
$\mathcal{E}_W=\{(\mathbf{e}_s,v_s)\}_{s=1}^{S}$, where
$\mathbf{e}_s\in\R^{d_e}$ and $v_s\in\{0,1\}$ indicates reference
availability. The target transcript is
\begin{equation}
\hat y_q=\operatorname*{arg\,max}_y
p_\theta(y\mid\mathbf{X},\mathbf{A},\mathcal{E}_W,q).
\label{eq:ssc_task}
\end{equation}

In Fig.~\ref{fig:ssc_framework}, RSC supplies the acoustic identity evidence
$\mathcal{E}_W$. TCRE combines this evidence with the activity tracks to
construct $\mathbf{c}_q(t)$, and CLRM maps the aligned condition to encoder
transformations. This computation is expressed as
\begin{align}
\mathcal{E}_W
  &=\mathcal{R}_{\mathrm{RSC}}(\mathbf{X}^{\mathrm{sess}},
       \mathbf{A}^{\mathrm{sess}};W), \label{eq:ssc_reference_map}\\
\mathbf{c}_q(t)
  &=\Phi_{\mathrm{TCRE}}(\mathbf{A},\mathcal{E}_W;q)(t),
       \label{eq:ssc_condition_map}\\
\bar{\mathbf{c}}_q
  &=\mathcal{P}_{\mathrm{enc}}(\mathbf{c}_q),
       \label{eq:ssc_alignment}\\
\widetilde{\mathbf{h}}_{\ell,\tau}
  &=\mathcal{A}_\ell(\bar{\mathbf{c}}_q(\tau))\mathbf{h}_{\ell,\tau}
       +\boldsymbol{\beta}_\ell(\bar{\mathbf{c}}_q(\tau)).
       \label{eq:ssc_affine}
\end{align}
Here $\mathcal{P}_{\mathrm{enc}}$ aligns frame-level conditions with the
backbone's acoustic tokens, $\tau$ indexes these tokens, and
$\mathbf{h}_{\ell,\tau}\in\R^D$ is the acoustic representation before
modulation at position $\ell$. CLRM specifies the matrix
$\mathcal{A}_\ell$ and offset $\boldsymbol{\beta}_\ell$. There is a
pre-stack position $\ell=0$ and one position before each of the $L$ encoder
blocks, $\ell=1,\ldots,L$. The resulting encoder features pass through the
backbone's native audio projection and decoder to generate $y_q$.

\subsection{Target-Centric Relational Encoding}
\label{sec:tcre}
TCRE determines which local activity information enters the target
condition. It encodes each track's history before aggregation, preserving
overlap and alternation patterns as target-centric relations. RSC identity
coordinates augment these track features, allowing the same relational
construction to incorporate acoustic evidence about who is speaking.

\subsubsection{From compressed states to track histories}
Let $\mathcal{O}_q$ be the set of valid non-target tracks and
$N_q=|\mathcal{O}_q|$. Padded tracks are excluded from this set. The
activity compression discussed in Section~\ref{sec:related_work} motivates
track-level encoding. In the STNO construction used by DiCoW \cite{polok2024target,polok2025dicow},
$u_q(t)=\prod_{s\in\mathcal{O}_q}(1-a_s(t))$ summarizes the non-target
tracks, giving the four coefficients
\begin{equation}
\begin{aligned}[b]
p_{q,\mathrm{S}}&=(1-a_q)u_q,
& p_{q,\mathrm{T}}&=a_qu_q,\\
p_{q,\mathrm{N}}&=(1-a_q)(1-u_q),
& p_{q,\mathrm{O}}&=a_q(1-u_q).
\end{aligned}
\label{eq:ssc_stno}
\end{equation}
For binary activities, these coefficients indicate silence, target-only,
non-target-only, and target-overlap states. Aggregating the non-target tracks
in this way merges different interferer counts and hand-over histories.
TCRE separately encodes tracks and computes symmetric
relations around the target.

A shared causal temporal encoder maps each activity track to
$\boldsymbol{\xi}_s(t)\in\R^d$. With $\operatorname{SiLU}$ activations,
the two convolutional layers are
\begin{align}
\boldsymbol{\zeta}_s(t)
 &=\operatorname{SiLU}\!\left(
     \sum_{j=0}^{4}\mathbf{w}_1[j]a_s(t-j)+\mathbf{b}_1\right),
     \label{eq:tcre_temporal_first}\\
\boldsymbol{\xi}_s(t)
 &=\operatorname{SiLU}\!\left(
     \sum_{j=0}^{4}\mathbf{W}_2[j]\boldsymbol{\zeta}_s(t-4j)
     +\mathbf{b}_2\right).
     \label{eq:tcre_temporal_second}
\end{align}
The kernels have size five and dilations one and four, respectively;
$\mathbf{w}_1[j]\in\R^d$, $\mathbf{W}_2[j]\in\R^{d\times d}$, and
both biases have dimension $d$. We use $d=32$. The receptive field covers
21 activity frames, or 210\,ms on the 10\,ms grid. Values outside the valid
window are zero padded, and invalid intermediate positions are masked.
Shared weights make the temporal encoding independent of track numbering.

The local feature and identity-augmented track feature are
\begin{align}
\mathbf{z}_s(t)&=[a_s(t);\boldsymbol{\xi}_s(t)]\in\R^{d+1},
       \label{eq:tcre_local}\\
\boldsymbol{\phi}_s(t)&=[1;\mathbf{z}_s(t);\boldsymbol{\eta}_s]
       \in\R^{d_\phi},\qquad d_\phi=d+d_e+3.
       \label{eq:tcre_track}
\end{align}
The vector $\boldsymbol{\eta}_s\in\R^{d_e+1}$ carries the reference
identity and its validity, as defined in Section~\ref{sec:rsc}. For a fixed
target request, it remains constant over time. The leading constant in
$\boldsymbol{\phi}_s$ permits the factorized relations below to include
constant and lower-order terms as well as interactions.

\subsubsection{Factorized second- and third-order relations}
TCRE represents the target's relation to individual interferers and to pairs
of distinct interferers.

For $k\in\{2,3\}$, let
$\mathbf{Q}_k,\mathbf{K}_k\in\R^{R\times d_\phi}$ be learned target-side
and non-target-side projections, with relation width $R=16$. Writing
$\mathbf{k}_s=\mathbf{K}_3\boldsymbol{\phi}_s(t)$ for the third-order
non-target factors, we compute
\begin{align}
\mathbf{g}_{2,q}(t)
 &=\big(\mathbf{Q}_2\boldsymbol{\phi}_q(t)\big)\odot
     \sum_{s\in\mathcal{O}_q}\mathbf{K}_2\boldsymbol{\phi}_s(t),
     \label{eq:tcre_second}\\
\mathbf{g}_{3,q}(t)
 &=\big(\mathbf{Q}_3\boldsymbol{\phi}_q(t)\big)\odot
     \sum_{\substack{i<j\\i,j\in\mathcal{O}_q}}
       \mathbf{k}_i\odot\mathbf{k}_j.
     \label{eq:tcre_third}
\end{align}
The second-order term relates the target to one non-target track; the
third-order term relates the target to two different non-target tracks.
These orders count the participating tracks. Since the track features
contain temporal history, a pairwise history interaction also describes
hand-overs between interferers that are not currently active together.

The pair sum is evaluated without enumerating all pairs:
\begin{equation}
\sum_{\substack{i<j\\i,j\in\mathcal{O}_q}}\mathbf{k}_i\odot\mathbf{k}_j
 =\frac12\left[
   \left(\sum_{i\in\mathcal{O}_q}\mathbf{k}_i\right)^{\odot2}
   -\sum_{i\in\mathcal{O}_q}\mathbf{k}_i^{\odot2}\right],
\label{eq:tcre_linear_pairs}
\end{equation}
where $\odot$ and $(\cdot)^{\odot2}$ denote element-wise multiplication and
squaring. This identity removes self-pairs, making aggregation linear
in $N_q$ for fixed feature dimensions. The third-order term is zero with
fewer than two valid non-target tracks.

\begin{figure}[!t]
\centering
\includegraphics[width=\columnwidth]{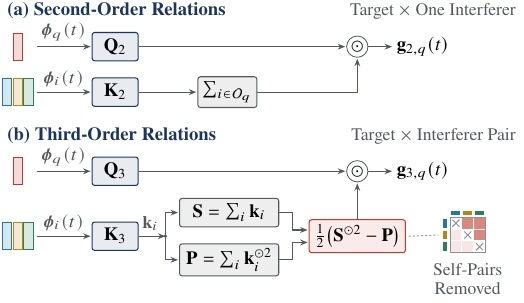}
\caption{Factorized target-centric relations. (a) The second-order branch
pools individual interferer factors. (b) The third-order branch combines
the target with distinct interferer pairs. Subtracting $\mathbf P$ from
$\mathbf S^{\odot2}$ removes self-pairs, and the factor $1/2$ counts each
unordered pair once. The grid illustrates this pair selection. Sums run
over valid non-target tracks $\mathcal O_q$, giving $O(N_q)$ pooling,
with $\mathbf k_i=\mathbf K_3\boldsymbol\phi_i(t)$ and element-wise
multiplication $\odot$.}
\label{fig:tcre_relations}
\end{figure}

Fig.~\ref{fig:tcre_relations} summarizes the two branches, including the
shared projections and self-pair removal in the third-order pool.

\subsubsection{Fixed-dimensional target condition and symmetry}
The complete frame-level condition is
\begin{equation}
\mathbf{c}_q(t)=
[\mathbf{z}_q(t);\mathbf{g}_{2,q}(t);\mathbf{g}_{3,q}(t);
 n_q(t);N_q;\boldsymbol{\eta}_q],
\label{eq:tcre_complete_condition}
\end{equation}
where $n_q(t)=\sum_{s\in\mathcal{O}_q}a_s(t)$ is the current non-target
activity mass. Thus $n_q(t)$ describes current activity, whereas $N_q$
counts valid non-target tracks in the window. With $d=d_e=32$ and $R=16$,
$d_\phi=67$ and $\mathbf{c}_q(t)\in\R^{100}$, independently of the number
of speakers. The activity-only model omits the 33 identity coordinates,
giving a 67-dimensional condition.

For binary activities, $u_q(t)=\mathbb{I}[n_q(t)=0]$, so the retained
coordinates $(a_q(t),n_q(t))$ determine the STNO coefficients before temporal
alignment. The additional history and relation coordinates express the finer
activity structures motivating TCRE: $n_q$ distinguishes one interferer
from two, and the temporal relations describe which tracks interfere over time.

For any permutation $\sigma$ of the speaker tracks, the condition satisfies
\begin{equation}
\Phi_{\mathrm{TCRE}}(\sigma\mathbf{A},\sigma\mathcal{E}_W;\sigma(q))
 =\Phi_{\mathrm{TCRE}}(\mathbf{A},\mathcal{E}_W;q).
\label{eq:tcre_equivariance}
\end{equation}
The temporal encoder and projection matrices are shared across tracks, and
all non-target aggregations are symmetric sums over tracks or unordered
pairs, following the shared-encoding and symmetric-aggregation principle
of set representations \cite{zaheer2017deepsets}. Consequently, relabeling
the tracks and requested target preserves
the condition for the same physical speaker. The target
conditions are thus permutation equivariant.

\subsection{Reference-Based Speaker Conditioning}
\label{sec:rsc}
\subsubsection{Diarization-guided session evidence}
\begin{figure}[!t]
\centering
\includegraphics[width=\columnwidth]{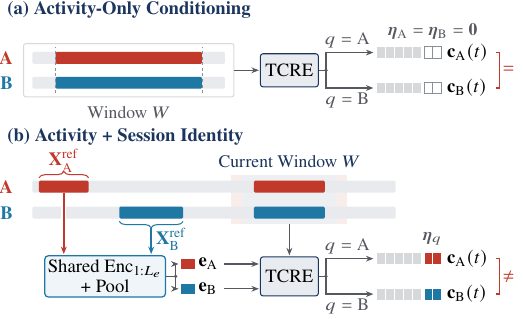}
\caption{Session identity distinguishes locally ambiguous targets.
(a) Identical local activity histories yield the same activity-only target
condition. (b) RSC encodes speaker-exclusive references outside the current
window $W$. For valid, distinct identities
($v_{\mathrm A}=v_{\mathrm B}=1$, $\mathbf e_{\mathrm A}\ne\mathbf e_{\mathrm B}$),
the target identity coordinates $\boldsymbol\eta_q$ distinguish the two
conditions. Colored cells highlight these coordinates. The timelines are
schematic.}
\label{fig:rsc_identity_evidence}
\end{figure}

Activity histories describe temporal interactions, but two targets with
identical activity throughout the available local history receive the same
activity-only TCRE condition. RSC addresses this ambiguity by supplying
session-level acoustic identity to the track features and the target
coordinates of $\mathbf{c}_q(t)$. This binds the requested track to a voice
within the relational condition.
Fig.~\ref{fig:rsc_identity_evidence} illustrates how acoustic identities
distinguish targets with identical local activity.
Algorithm~\ref{alg:rsc_condition} summarizes reference selection and
condition construction.

For each speaker, the selection map
\begin{equation}
(\mathbf{X}^{\mathrm{ref}}_s,v_s)
 =\mathcal{S}_s(\mathbf{X}^{\mathrm{sess}},\mathbf{A}^{\mathrm{sess}};W)
\label{eq:rsc_selection}
\end{equation}
collects speaker-exclusive evidence from other intervals of the session.
We exclude the current window with a temporal margin, subtract
intervals assigned to other speakers, and trim candidate boundaries.
Among segments meeting a minimum segment duration, we select up to
$K_{\mathrm{ref}}$, prioritizing longer segments and capping the total
duration at $T_{\max}$.
Distance to $W$ and start time resolve ties. A track has $v_s=1$ when
at least $T_{\min}$ of selected evidence is available and $v_s=0$ otherwise.
At inference, input diarization supplies the
selection intervals. Hyperparameter values are given in
Section~\ref{sec:experimental_setup}.

For online simulated mixtures, references are independent clean utterances
of the same source identity. The selection excludes the utterances used in
the current mixture and candidates with the same normalized text. This
provides acoustic evidence for identity across different speech content.

\subsubsection{Shared reference encoding and identity pooling}
Reference audio passes through the shared acoustic encoder prefix with
target-only activity conditioning. We use the first $L_e=6$ encoder blocks and
pool their valid output tokens. Let
$\mathbf{h}^{\mathrm{ref}}_{s,\tau}\in\R^D$ denote these representations
and $\mathbf{r}_{s,\tau}=\operatorname{LN}
(\mathbf{h}^{\mathrm{ref}}_{s,\tau})$. Using learned token weights as in
attentive speaker pooling \cite{okabe2018attentive}, we form a weighted mean:
\begin{align}
\alpha_{s,\tau}
 &=\frac{\exp(\mathbf{w}_\alpha^\top\mathbf{r}_{s,\tau})}
 {\sum_{\tau'}\exp(\mathbf{w}_\alpha^\top\mathbf{r}_{s,\tau'})},
       \label{eq:rsc_attention}\\
\mathbf{e}_s
 &=\mathbf{W}_e\sum_\tau\alpha_{s,\tau}\mathbf{r}_{s,\tau}
       +\mathbf{b}_e\in\R^{d_e}.
       \label{eq:rsc_identity}
\end{align}
Here $\mathbf{w}_\alpha\in\R^D$,
$\mathbf{W}_e\in\R^{d_e\times D}$, and $d_e=32$.
The sums include only valid reference tokens. Reference encoding and pooling
are differentiable under the target-transcription loss.

For a request targeting $q$, the identity features are gated by target
reference availability:
\begin{equation}
\widetilde v_{s\mid q}=v_sv_q,\qquad
\boldsymbol{\eta}_s=
[\widetilde v_{s\mid q}\mathbf{e}_s;\widetilde v_{s\mid q}].
\label{eq:rsc_validity_gate}
\end{equation}
We set unavailable identities to zero. If $v_q=0$, all
$\boldsymbol{\eta}_s$ are zero and the request uses activity-only
conditioning. With a valid target reference, each available non-target
identity participates in the relational features. The target identity also
enters $\mathbf{c}_q$ directly in
\eqref{eq:tcre_complete_condition}. Identity therefore influences both the
relations and the final condition supplied to the modulation generator.

\begin{figure}[!t]
\centering
\begin{minipage}{\columnwidth}
\small
\refstepcounter{algorithm}\label{alg:rsc_condition}
\hrule height 0.8pt
\vspace{3pt}
\noindent\textbf{Algorithm~\thealgorithm\quad RSC condition construction}
\par\vspace{3pt}\hrule height 0.4pt\vspace{4pt}
\noindent\textbf{Input:} Session mixture $\mathbf{X}^{\mathrm{sess}}$ and
diarization $\mathbf{A}^{\mathrm{sess}}$; local activity $\mathbf{A}$,
current window $W$, target $q$, and reference hyperparameters
$K_{\mathrm{ref}},T_{\min},T_{\max}$.
\par\noindent\textbf{Output:} Frame-level target condition $\mathbf{c}_q(t)$.
\par\vspace{4pt}
\setlength{\tabcolsep}{0pt}
\renewcommand{\arraystretch}{1.03}
\begin{tabular}{@{}r@{\hspace{5pt}}>{\raggedright\arraybackslash}p{\dimexpr\linewidth-15pt\relax}@{}}
1 & \textbf{for} $s=1,\ldots,S$ \textbf{do} \\
2 & \hspace*{0.6em}Select speaker-exclusive evidence $\mathbf{X}^{\mathrm{ref}}_s$
    outside $W$ using \eqref{eq:rsc_selection}. \\
3 & \hspace*{0.6em}Keep up to $K_{\mathrm{ref}}$ segments, totaling at most $T_{\max}$. \\
4 & \hspace*{0.6em}Set $v_s\leftarrow1$ if total duration $\ge T_{\min}$,
    else $v_s\leftarrow0$. \\
5 & \hspace*{0.6em}Compute $\mathbf{e}_s$ by \eqref{eq:rsc_identity} if $v_s=1$;
    otherwise set $\mathbf{e}_s=\mathbf{0}$. \\
6 & \textbf{end for} \\
7 & Use $\widetilde v_{s\mid q}=v_s v_q$ in $\boldsymbol{\eta}_s$
    for all $s$ \eqref{eq:rsc_validity_gate}. \\
8 & $\mathcal{E}_W\leftarrow\{(\mathbf{e}_s,v_s)\}_{s=1}^{S}$;
    $\mathbf{c}_q(t)\leftarrow\Phi_{\mathrm{TCRE}}(\mathbf{A},\mathcal{E}_W;q)(t)$. \\
9 & \textbf{return} $\mathbf{c}_q(t)$ \\
\end{tabular}
\par\vspace{4pt}
\noindent\textbf{Training:} Sample real / simulated / binding views with
weights $\boldsymbol{\rho}$. For a binding view, use two targets from the same mixture
with matched durations and minimize $\mathcal{L}_{\mathrm{bind}}$
in \eqref{eq:rsc_binding_loss}.
\par\vspace{4pt}\hrule height 0.8pt
\end{minipage}
\end{figure}

\subsubsection{Paired target binding}
Training includes paired requests for the same highly overlapped mixture.
Two targets A and B have matched speech durations satisfying
$\min(\delta_{\mathrm A},\delta_{\mathrm B})/
\max(\delta_{\mathrm A},\delta_{\mathrm B})\ge0.95$.
Both views share the mixture and augmentation randomness and are kept
together within a device batch. Their supervised contribution is
\begin{equation}
\mathcal{L}_{\mathrm{bind}}
 =-\sum_{s\in\{\mathrm A,\mathrm B\}}
   \log p_\theta(y_s\mid\mathbf{X},\bar{\mathbf{c}}_s).
\label{eq:rsc_binding_loss}
\end{equation}
Paired requests train target selection using acoustic identity when activity
patterns provide little separation. Real, ordinary simulated, and
paired-binding views use sampling weights
$\boldsymbol{\rho}=(\rho_{\mathrm{real}},\rho_{\mathrm{sim}},\rho_{\mathrm{bind}})$
(Section~\ref{sec:experimental_setup}).

\subsection{Conditional Low-Rank Modulation}
\label{sec:clrm}
The joint TCRE--RSC condition specifies the requested speaker and the
surrounding activity structure. CLRM makes this information control
recognition by generating both channel-wise gains and offsets and
condition-dependent combinations of acoustic channels.
CLRM combines feature-wise affine conditioning~\cite{perez2018film}
with a compact cross-channel transform, using a hypernetwork to generate
condition-dependent parameters~\cite{ha2017hypernetworks}.
Low-rank-plus-diagonal acoustic adaptation~\cite{zhao2016lrpd} decomposes
speaker transforms into diagonal and low-rank terms; bottleneck
adapters~\cite{houlsby2019adapters} and LoRA~\cite{hu2022lora} use compact
modules or weight updates for efficient adaptation.
In CLRM, learned read/write bases are shared across conditions, while the
diagonal parameters and a small mixing core vary with the target's activity
and identity at each encoder token.

\subsubsection{Alignment with the acoustic token grid}
Relations are constructed at the activity-frame rate and then aligned with
the backbone's convolutional subsampling. For stride $\kappa$, define the
valid-normalized pooling operator
\begin{equation}
\mathcal{P}_\kappa(\mathbf{c})(\tau)
 =\frac{\sum_{j=-1}^{1}m(\kappa\tau+j)\mathbf{c}(\kappa\tau+j)}
 {\max\!\left(1,\sum_{j=-1}^{1}m(\kappa\tau+j)\right)},
\label{eq:clrm_valid_pool}
\end{equation}
where $m$ marks valid input positions. Valid output lengths and masks are
recomputed after each stage, separately for each acoustic chunk. For
Qwen3-ASR,
$\mathcal{P}_{\mathrm{enc}}=\mathcal{P}_2\circ\mathcal{P}_2\circ
\mathcal{P}_2$, mapping 10\,ms frames to 80\,ms tokens. FireRedAudio
uses $\mathcal{P}_2\circ\mathcal{P}_1$ to match its native encoder grid. Validity normalization prevents padding from diluting the
condition. The signed relation coordinates are averaged as features.

\subsubsection{Shared generator and position-specific heads}
All modulation positions use a shared hidden layer, with a learned
position embedding $\boldsymbol{\omega}_\ell\in\R^{d_\omega}$:
\begin{equation}
\boldsymbol{\psi}_{\ell,\tau}
 =\operatorname{SiLU}\!\left(
 \mathbf{W}_g[\bar{\mathbf{c}}_q(\tau);
 \boldsymbol{\omega}_\ell]+\mathbf{b}_g\right)\in\R^{d_h}.
\label{eq:clrm_generator}
\end{equation}
We use $d_h=32$ and $d_\omega=8$. Each position has an affine output head,
\begin{equation}
[\boldsymbol{\gamma}_{\ell,\tau};\boldsymbol{\beta}_{\ell,\tau}]
 =\mathbf{W}^{\mathrm f}_\ell\boldsymbol{\psi}_{\ell,\tau}
  +\mathbf{b}^{\mathrm f}_\ell\in\R^{2D},
\label{eq:clrm_film_head}
\end{equation}
and each encoder-block position has a low-rank core head,
\begin{equation}
\mathbf{M}_{\ell,\tau}
 =\frac{1}{\sqrt r}\operatorname{reshape}_{r\times r}\!\left[
 \tanh\!\left(\mathbf{W}^{\mathrm x}_\ell\boldsymbol{\psi}_{\ell,\tau}
 +\mathbf{b}^{\mathrm x}_\ell\right)\right].
\label{eq:clrm_core_head}
\end{equation}
The head dimensions are
$\mathbf{W}^{\mathrm f}_\ell\in\R^{2D\times d_h}$ and
$\mathbf{W}^{\mathrm x}_\ell\in\R^{r^2\times d_h}$.
Thus the same activity--identity condition controls channel-wise gains,
offsets, and cross-channel mixing. The core rank is $r=8$, distinct from the
TCRE relation width $R=16$.

\subsubsection{Parallel acoustic modulation}
Let $\mathbf{U}_\ell,\mathbf{V}_\ell\in\R^{D\times r}$ be learned
write and read bases shared across conditions at position $\ell$. The
operator in \eqref{eq:ssc_affine} is
\begin{equation}
\mathcal{A}_\ell(\mathbf{c})
 =\operatorname{Diag}(\mathbf{1}+\boldsymbol{\gamma}_\ell(\mathbf{c}))
  +\mathbf{U}_\ell\mathbf{M}_\ell(\mathbf{c})\mathbf{V}_\ell^\top.
\label{eq:clrm_operator}
\end{equation}
The first term provides FiLM \cite{perez2018film}; the second, referred to as
the cross-channel low-rank (XLR) branch, reads the
acoustic feature into an $r$-dimensional space, mixes it with the generated
core, and writes the result back. Both terms act in parallel on the same
pre-modulation representation $\mathbf{h}_{\ell,\tau}$. The pre-stack
position uses $\mathbf{M}_0\equiv\mathbf{0}$; the cross-channel term is
applied before each encoder block.

Writing $\mathbf{u}_{\ell,i}$ and $\mathbf{v}_{\ell,j}$ for columns of
the two bases makes the conditional channel combinations explicit:
\begin{equation}
\mathbf{U}_\ell\mathbf{M}_\ell(\mathbf{c})\mathbf{V}_\ell^\top
 =\sum_{i,j=1}^{r}[\mathbf{M}_\ell(\mathbf{c})]_{ij}
   \mathbf{u}_{\ell,i}\mathbf{v}_{\ell,j}^\top.
\label{eq:clrm_basis}
\end{equation}
The coefficients of these read/write combinations vary with the target
condition. The additional matrix has rank at most $r$ and costs
$O(Dr+r^2)$ to apply per token. Each block adds
$2Dr+(d_h+1)r^2$ parameters for the bases and core head. Setting
$\mathbf{M}_\ell=0$ yields diagonal FiLM modulation.

The affine and core output heads are zero initialized, giving an identity
transform at the start of fresh training. The read/write bases are
orthogonally initialized. Condition construction, parameter generation, and
the low-rank branch use FP32 arithmetic. The resulting acoustic updates
are converted to match the numerical precision of the backbone's hidden
representations. For Qwen3-ASR-0.6B ($D=896$, $L=18$),
the cross-channel branches add 296,064 parameters. Complete SSC, including
TCRE, RSC pooling, and CLRM, contains 1,464,312 trainable conditioning parameters.

\subsection{Training and Target-Conditioned Inference}
\label{sec:ssc_training}
The supervised objective uses only the requested target's transcript. For
$K_q$ output tokens, it is
\begin{equation}
\mathcal{L}_{\mathrm{SFT}}
 =-\sum_{k=1}^{K_q}\log p_\theta
 (y_{q,k}\mid y_{q,<k},\mathbf{X},\bar{\mathbf{c}}_q).
\label{eq:ssc_sft}
\end{equation}
All SSC modules, including the reference pooling path, learn through this
objective. Binding pairs contribute the two target-specific losses in
\eqref{eq:rsc_binding_loss}.

After supervised fine-tuning, GSPO \cite{zheng2025gspo} optimizes target
transcription error. For each target-conditioned input $C_q$, the sampling
policy $\pi_{\mathrm{old}}$ generates a group of $G$ candidates. Candidate
$i$ receives reward
\begin{equation}
R_i=-\frac{\mathrm{S}_i+\mathrm{D}_i+\mathrm{I}_i}
 {\max(N_q^{\mathrm{ref}},1)}
 -\lambda_{\mathrm{cap}}\mathbb{I}[\text{candidate $i$ is truncated}],
\label{eq:ssc_reward}
\end{equation}
where edit counts are computed after text normalization against the requested
target's window transcript, using characters for Chinese and words for English. Let
$\widehat A_i=(R_i-\bar R)/(\sigma_R+\epsilon_A)$ denote the group-normalized
advantage, where $\bar R$ and $\sigma_R$ are the group mean and population
standard deviation. With candidate length $K_i$, the sequence-level importance ratio
and objective are
\begin{align}
w_i(\theta)
 &=\left(\frac{\pi_\theta(\hat y^{(i)}\mid C_q)}
 {\pi_{\mathrm{old}}(\hat y^{(i)}\mid C_q)}\right)^{1/K_i},
       \label{eq:ssc_gspo_ratio}\\
\mathcal{L}_{\mathrm{GSPO}}
 &=-\frac1G\sum_{i=1}^{G}
 \min\!\left(w_i\widehat A_i,
       \operatorname{clip}(w_i,1-\varepsilon_-,1+\varepsilon_+)
       \widehat A_i\right)\nonumber\\
 &\quad+\lambda_{\mathrm{CE}}\mathcal{L}_{\mathrm{CE}}.
       \label{eq:ssc_gspo_loss}
\end{align}
Here $\mathcal{L}_{\mathrm{CE}}$ is the token-normalized supervised loss
across the logical minibatch. The text-system recipe uses $G=8$, $\lambda_{\mathrm{cap}}=1$,
$\epsilon_A=10^{-4}$, $\varepsilon_-=3\times10^{-4}$,
$\varepsilon_+=4\times10^{-4}$, and $\lambda_{\mathrm{CE}}=0.01$.
Target identities and activity conditions accompany each
candidate during sampling and optimization.

At inference, a long session is divided into windows. For each requested
track, RSC obtains session evidence, TCRE constructs its local condition,
and CLRM modulates the encoder used to decode that target. Window outputs
are assembled using the session speaker identities. A timestamp-output
variant represents $y_q$ as sentences with window-relative onset and offset
times, with the same conditioning modules.

\section{Experiments}
\label{sec:experiments}
\subsection{Experimental Setup}
\label{sec:experimental_setup}
\subsubsection{Data and speaker priors}
Table~\ref{tab:sets} summarizes the data. Data preparation builds on
mt-asr-data-prep.\footnote{\url{https://github.com/BUTSpeechFIT/mt-asr-data-prep}}
Real meeting training uses the
training partitions of AliMeeting~\cite{yu2022m2met},
AISHELL-4~\cite{fu2021aishell4}, AMI~\cite{carletta2006ami}, and
NOTSOFAR-1~\cite{vinnikov2024notsofar}, totaling approximately 331~h of
single-channel recordings. Qwen3-ASR and FireRedAudio additionally use a
10,000-h TTS source pool to construct online 30-s mixtures with two to six
speakers and 0--60\% overlap, including room-response augmentation.
Real and simulated target views are sampled at a 2:1 ratio. With RSC,
the sampling weights for real, ordinary simulated, and paired-binding
views are $\boldsymbol{\rho}=(20,9,1)$.
Reference selection uses $K_{\mathrm{ref}}=4$, $T_{\min}=2\,\mathrm{s}$,
and $T_{\max}=8\,\mathrm{s}$, with a 0.5-s exclusion margin,
0.2-s boundary trimming, and a 0.5-s minimum segment duration.

\begin{table}[!t]
\caption{Training and Evaluation Data. SDM: single distant microphone; SC: single-channel device recording}
\label{tab:sets}
\centering\footnotesize
\setlength{\tabcolsep}{4pt}
\renewcommand{\arraystretch}{1.0}
\sbox{\TableBody}{%
\begin{tabular*}{\columnwidth}{@{\extracolsep{\fill}}llcr@{}}
\toprule
Dataset & Partition / channel & Lang. & Hours \\
\midrule
\multicolumn{4}{@{}l}{\emph{Real meeting training data}}\\
AliMeeting & Train, far-field ch.\,0 & zh & 104.75 \\
AISHELL-4 & Train, far-field ch.\,0 & zh & 107.50 \\
AMI-SDM & Train, SDM & en & $\approx$81 \\
NOTSOFAR-1 & Train 1+2, SC & en & $\approx$38 \\
\midrule
\multicolumn{4}{@{}l}{\emph{Auxiliary augmentation}}\\
TTS source pool & Online mixtures & zh/en & 10k \\
\midrule
\multicolumn{4}{@{}l}{\emph{In-domain evaluation}}\\
AliMeeting & Test, far-field ch.\,0 & zh & 10.00 \\
AISHELL-4 & Test, far-field ch.\,0 & zh & 12.72 \\
AMI-SDM & Eval, SDM & en & $\approx$9 \\
NOTSOFAR-1 & Eval-small, SC & en & 16.00 \\
\midrule
\multicolumn{4}{@{}l}{\emph{Out-of-domain evaluation}}\\
MISP-Meeting & Eval, far-field & zh & 3.00 \\
AISHELL-5 & Eval2, far-field & zh & 3.58 \\
DiPCo & Eval, far-field & en & $\approx$2.6 \\
\bottomrule
\end{tabular*}%
}
\usebox{\TableBody}
\end{table}

The four in-domain test sets contain 20 AliMeeting recordings, 20
AISHELL-4 recordings, 16 AMI-SDM recordings, and 160 NOTSOFAR-1 device
recordings from 80 meetings. All systems use far-field single-channel audio:
channel~0 for AliMeeting and AISHELL-4, SDM with the Full-corpus-ASR
partition for AMI, and the single-channel
device tracks for NOTSOFAR-1. Out-of-domain evaluation uses six
MISP-Meeting recordings~\cite{gao2025misp}, 18 AISHELL-5 Eval2
recordings~\cite{dai2025aishell5}, and five DiPCo Eval
recordings~\cite{vansegbroeck2019dipco}, with no training or adaptation
on these evaluation sets.

Mechanism ablations primarily use oracle diarization to isolate the
conditioning design from input diarization errors. Unless otherwise
specified, predicted priors are obtained by DiariZen~\cite{han2025diarizen}
with a W2V-BERT2 frontend, followed by
sequence-to-sequence (S2S) refinement~\cite{wang2026s2srefinement}.
RSC selects its inference reference segments from this input prior.
At a 0.25-s collar, input DERs for AliMeeting, AISHELL-4, AMI-SDM, and
NOTSOFAR-1 are 3.62/4.89/8.65/5.10\%.

\begin{table*}[!t]
\caption{Component Ablations on Qwen3-ASR-0.6B with Oracle Diarization. $\mathbf{c}^{(\leq K)}$ uses activity relations through order $K$, and $\mathcal E$ adds RSC identities. Par. gives conditioning parameters in millions. RER is relative to STNO-FDDT. The 1/2/$\geq$3-spk columns group windows by peak reference concurrency. Bold and underline mark the best and second-best values before rounding}
\label{tab:ladder}
\centering\footnotesize
\setlength{\tabcolsep}{2.1pt}
\renewcommand{\arraystretch}{1.05}
\sbox{\TableBody}{%
\begin{tabular*}{\textwidth}{@{\extracolsep{\fill}}llcccccccccc@{}}
\toprule
\multirow{2}{*}{Cond.\ $\mathbf{c}$} & \multirow{2}{*}{Operator} & \multirow{2}{*}{Par.\,(M)} & \multicolumn{4}{c}{Corpus cpER (\%)$\downarrow$} & \multicolumn{5}{c}{RER vs.\ STNO-FDDT (\%)$\uparrow$} \\
\cmidrule(lr){4-7}\cmidrule(lr){8-12}
& & & AliMeeting & AISHELL-4 & AMI-SDM & NOTSOFAR-1 & All & 1 spk & 2 spk & $\geq$3 spk & No-solo target \\
\midrule
STNO & FDDT & 0.14 & 17.92 & 13.20 & 18.35 & 21.36 & -- & -- & -- & -- & -- \\ \midrule
STNO & FiLM & 1.13 & 17.68 & 13.23 & 18.12 & 20.90 & 1.14 & 0.10 & 2.21 & 0.73 & 0.13 \\
$\mathbf{c}^{(\leq 2)}$ & FiLM & 1.13 & 18.03 & 13.07 & 18.31 & 20.97 & 0.62 & 2.33 & $-0.15$ & 0.19 & 0.97 \\
$\mathbf{c}^{(\leq 3)}$ & FiLM & 1.13 & 17.18 & 12.95 & 18.15 & 20.47 & 2.84 & \textbf{5.70} & 1.03 & 3.42 & 2.65 \\ \midrule
$\mathbf{c}^{(\leq 3)}$ & FiLM$_{64}$ & 2.23 & 17.76 & 13.40 & 18.14 & 20.68 & 0.92 & 0.03 & 0.19 & 1.97 & \underline{3.47} \\
$\mathbf{c}^{(\leq 3)}$ & XLR$_8$ & 0.31 & 17.57 & \underline{12.86} & 17.95 & 20.56 & 2.60 & \underline{5.36} & 1.67 & 2.83 & 1.25 \\
$\mathbf{c}^{(\leq 3)}$ & CLRM$_8$ & 1.43 & \underline{16.96} & 12.92 & \underline{17.81} & \underline{20.27} & \underline{3.90} & 5.33 & \underline{2.48} & \underline{4.49} & 2.66 \\ \midrule
$\mathbf{c}^{(\leq3)}(\mathcal{E})$ & CLRM$_8$ & 1.46 & \textbf{16.47} & \textbf{12.86} & \textbf{17.19} & \textbf{19.67} & \textbf{6.23} & 2.38 & \textbf{3.44} & \textbf{7.76} & \textbf{12.49} \\
\bottomrule
\end{tabular*}%
}
\usebox{\TableBody}
\end{table*}

\subsubsection{Models and optimization}
Our implementation extends TS-ASR-Whisper.\footnote{\url{https://github.com/BUTSpeechFIT/TS-ASR-Whisper}}
Our backbones are Qwen3-ASR-0.6B~\cite{qwen2026asr} and the 9B
FireRedAudio model~\cite{shen2026fireredaudio}.
Unless stated otherwise, the conditioning modules use the widths introduced
in Section~\ref{sec:method} ($d=d_e=d_h=32$, $R=16$, $r=8$).
The reference path shares the first six encoder blocks with the
recognition path. For both Qwen3-ASR and FireRedAudio, we jointly train
the acoustic encoder, audio projector, and conditioning modules, with rank-8 attention
LoRA~\cite{hu2022lora} in the decoder (scaling 16, dropout 0.05).
SFT uses eight GPUs, BF16 backbone
computation, and FP32
conditioning modules. The base learning rate is $4\times10^{-5}$;
conditioning parameters use a tenfold multiplier. Qwen3-ASR-0.6B uses a
15,000-step cosine schedule with 500 warmup steps; the activity-only
variants stop after 10,000 steps. FireRedAudio uses 30,000 steps
and 1,000 warmup steps.
For SFT and GSPO, we select checkpoints by minimizing the equally weighted
mean of autoregressive decoding error rates on the AliMeeting and AMI
development sets.

\subsubsection{Metrics and comparison protocol}
We report concatenated minimum-permutation error rate (cpER), using
characters for Chinese and words for English. After optimal speaker
permutation within each session, corpus-level scores are
\begin{equation}
\operatorname{cpER}=100\frac{\sum_m(S_m+D_m+I_m)}{\sum_m N_m}.
\label{eq:cp_micro}
\end{equation}
Here $S_m$, $D_m$, $I_m$, and $N_m$ denote substitutions, deletions,
insertions, and reference units in session $m$. Relative error reduction
(RER) is $100(e_{\mathrm{base}}-e_{\mathrm{new}})/e_{\mathrm{base}}$.
Multi-dataset means weight each dataset equally. RSC interventions use
fixed-target CER/WER, keeping the requested speaker assignment.

For timestamp outputs, MeetEval~\cite{neumann2023meeteval} computes
time-constrained cpER (tcpER) with a 5-s collar using character-based
pseudo-word timings. Diarization error rate (DER) uses a 0.25-s collar,
includes overlapping speech, and pools errors over scored speaker time.

Paired comparisons share reference transcripts and text normalization,
and complete-system comparisons use a common normalization.
For corpus-level comparisons, 95\% confidence intervals use 10,000 paired
meeting-bootstrap samples~\cite{bisani2004bootstrap}, with NOTSOFAR-1
device pairs resampled together.

\subsection{TCRE: Preserving Multi-Speaker Activity Relations}
\label{sec:tcre_experiment}
\begin{figure}[!t]
\centering
\includegraphics[width=\columnwidth]{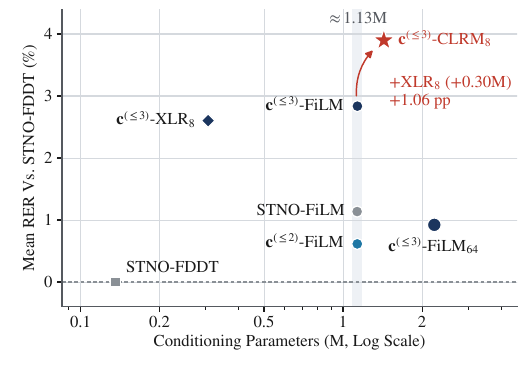}
\caption{Parameter efficiency of the activity-only ablations in
Table~\ref{tab:ladder}. Points show conditioning-module size and mean RER
over STNO-FDDT across the four test sets.}
\label{fig:parameter_efficiency}
\end{figure}

The configurations in Table~\ref{tab:ladder} progress from four-state
conditioning to TCRE, then to TCRE--CLRM, and finally to the full RSC
recipe. We first vary the activity representation while fixing the FiLM
operator and generator width. STNO-FiLM
encodes the four state trajectories with causal convolutions and concatenates
their temporal features with the state probabilities before FiLM parameter
generation. It provides a 1.14\% mean RER over STNO-FDDT. Second-order TCRE alone
gives 0.62\%, whereas adding the third-order term raises the reduction
to 2.84\%, with cpER 17.18/12.95/18.15/20.47\% across the four
sets. Thus, track-level conditioning is most effective when it includes
distinct-interferer-pair relations.
Fig.~\ref{fig:parameter_efficiency} shows that third-order TCRE improves
over STNO-FiLM at a similar conditioning budget of about 1.13M parameters.

We examine this relation using the original 30-s windows, grouped by
peak reference concurrency. The peak-1/2/$\geq$3 groups contain
1,221/2,022/2,660 windows. Errors are attributed after full-session
speaker matching and edit alignment: reference-unit midpoints locate
substitutions and deletions, and decoder windows locate insertions.
The third-order model improves over the second-order model by 3.22\%
in the peak-$\geq$3 group (95\% CI [1.85, 4.50]\%). Peak concurrency of
at least three supplies simultaneous non-target activity for the
distinct-interferer-pair term in \eqref{eq:tcre_third}. The third-order model
also improves over the second-order model by 1.14\% in the peak-2 group.
In peak-1 windows, the third- and second-order models achieve 5.70\%
and 2.33\% RER over STNO-FDDT, respectively (Table~\ref{tab:ladder}).

\subsection{CLRM: Conditional Cross-Channel Modulation}
\label{sec:clrm_experiment}
With the third-order TCRE condition fixed, we next vary how it controls
the acoustic encoder. Rank-8 CLRM improves the mean RER over STNO-FDDT
from FiLM's 2.84\% to 3.90\%
(Table~\ref{tab:ladder}). Increasing the FiLM generator width from
32 to 64 instead gives 0.92\% RER with 2.23M parameters,
compared with CLRM's 1.43M. The cross-channel-only variant is
compact (0.31M parameters) and gives 2.60\% RER; combining it
with the diagonal branch produces the strongest activity-only result.
These comparisons support combining diagonal
adaptation with low-rank channel mixing.

\begin{figure}[!t]
\centering
\includegraphics[width=\columnwidth]{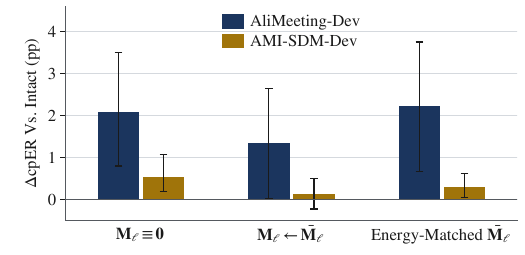}
\caption{Recognition results from frozen-CLRM-core interventions. Values are
changes in \emph{meeting-macro} cpER from the intact model on eight
AliMeeting-dev and four AMI-SDM-dev recordings, with 95\% paired
meeting-bootstrap CIs. The shared projections and all other weights
remain fixed. The mean-core interventions replace the varying core
with its calibration mean, optionally matching its energy.}
\label{fig:core_probe}
\end{figure}

We further intervene on the core of a frozen Qwen3-ASR-0.6B TCRE--CLRM
checkpoint (Fig.~\ref{fig:core_probe}). Setting the core to zero
increases meeting-macro cpER by 2.09~pp on AliMeeting-dev
(95\% CI [0.80, 3.51]) and 0.53~pp on AMI-SDM-dev
([0.19, 1.08]). Replacing it with a condition-independent calibration
mean also increases error; with energy matching, the increases are
2.22 and 0.30~pp, respectively. The trained model therefore uses
the varying core, beyond retaining a fixed cross-channel projection.
These interventions support the role of the condition-dependent core
described in Section~\ref{sec:clrm}.

\subsection{RSC: Anchoring the Requested Speaker}
\label{sec:rsc_experiment}
The full RSC recipe adds session identities and paired target binding
to TCRE--CLRM, yielding complete SSC. It achieves the best corpus scores in
Table~\ref{tab:ladder}, with cpERs of 16.47/12.86/17.19/19.67\%,
1.46M conditioning parameters, and a 6.23\% mean RER over STNO-FDDT.
Complete SSC also achieves 3.44\% RER over STNO-FDDT in the peak-2 group
and 7.76\% in the peak-$\geq$3 group.
The largest gain from the full RSC recipe occurs when the target has no solo
speech, i.e., every target frame overlaps with another speaker.
Relative to activity-only TCRE--CLRM, the full RSC recipe reduces error in
this group by 10.09\% (95\% CI [7.58, 12.09]\%); the
reduction is 0.86\% when more than half of the target's speech is solo.
Relative to STNO-FDDT, complete SSC reaches 12.49\% RER in this
fully overlapped group. The full RSC recipe is especially effective when
local activity poorly distinguishes targets.

\begin{figure}[!t]
\centering
\includegraphics[width=\columnwidth]{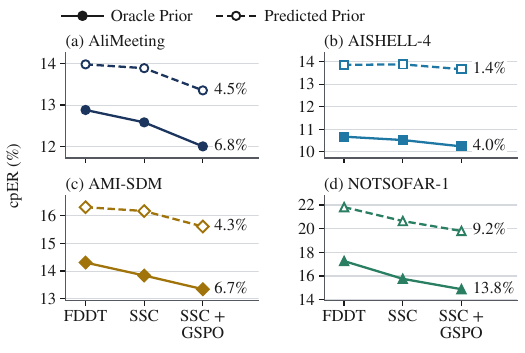}
\caption{FireRedAudio progression from FDDT to SSC and SSC + GSPO
on the four datasets (a)--(d), with oracle priors and predicted S2S-refined
priors. Endpoint labels give the relative cpER reduction over FDDT.}
\label{fig:fa_progression}
\end{figure}

Table~\ref{tab:rsc_evidence} measures the inference-time contribution of
session identity at a fixed FireRedAudio timestamp checkpoint after GSPO
(Section~\ref{sec:timestamp_experiment}).
With audio, target requests, reference candidates, and weights
fixed, we supply or remove RSC identities under the same local activity.
Starting from oracle tracks, we apply target misses, false alarms, and
local track swaps to 0.5-s activity blocks at a 10\% rate (seed~42).

\begin{table}[!t]
\caption{RSC under Activity Interventions. Fixed-target CER/WER (\%) without (w/o) and with (w/) RSC, and relative error reduction (RER). Bold marks lower error. The example uses identical intervened activity at the same checkpoint}
\label{tab:rsc_evidence}
\centering\footnotesize
\setlength{\tabcolsep}{1.6pt}
\renewcommand{\arraystretch}{1.0}
\sbox{\TableBody}{\parbox{\columnwidth}{%
\begin{tabular*}{\columnwidth}{@{}l@{\extracolsep{\fill}}rrrrrr@{}}
\toprule
\multirow{2}{*}{Activity condition} & \multicolumn{3}{c}{AliMeeting (CER)} & \multicolumn{3}{c}{AMI-SDM (WER)} \\
\cmidrule(lr){2-4}\cmidrule(lr){5-7}
& w/o & w/ & RER & w/o & w/ & RER \\
\midrule
Clean activity & 12.60 & \textbf{12.14} & 3.67 & 14.23 & \textbf{13.59} & 4.50 \\
Target miss (10\%) & 18.92 & \textbf{15.55} & 17.81 & 18.98 & \textbf{16.22} & 14.51 \\
Target-track swap (10\%) & 18.01 & \textbf{14.80} & 17.84 & 18.50 & \textbf{16.09} & 13.06 \\
Target false alarm (10\%) & \textbf{16.14} & 16.16 & $-0.12$ & \textbf{16.89} & 17.26 & $-2.16$ \\
\midrule
\multicolumn{7}{@{}l}{\textit{AMI example. ES2004d / MEO015, 1650--1680\,s}} \\
\end{tabular*}\par\vspace{1pt}
\begin{tabular*}{\columnwidth}{@{}l@{\extracolsep{\fill}}p{\dimexpr\columnwidth-1.55cm\relax}@{}}
Reference & seems fine except \textbf{voice} \textbf{recognition} everything \textbf{is} yeah \\
w/o RSC & seems fine except \textbf{for} \textbf{iso} \textbf{admission} everything \textbf{looks} yeah \\
w/ RSC & seems fine except voice recognition everything is yeah \\
\bottomrule
\end{tabular*}}%
}
\usebox{\TableBody}
\end{table}

\begin{table}[!t]
\caption{Timestamp Prediction with SSC and SSC + GSPO (\%). $\Delta$: (SSC + GSPO) $-$ SSC in cpER percentage points; $^{\dagger}$: 95\% paired meeting-bootstrap confidence interval below zero}
\label{tab:ts_gspo}
\centering\footnotesize
\setlength{\tabcolsep}{2.1pt}
\renewcommand{\arraystretch}{1.0}
\sbox{\TableBody}{%
\begin{tabular}{@{}llccccc@{}}
\toprule
& & \multicolumn{3}{c}{cpER} & \multicolumn{2}{c}{tcpER 5\,s} \\
\cmidrule(lr){3-5}\cmidrule(l){6-7}
Prior & Set & SSC & + GSPO & $\Delta$ & SSC & + GSPO \\
\midrule
\multirow{4}{*}{Oracle} & AliMeeting & 12.56 & \textbf{12.11} & $-0.45^{\dagger}$ & 12.65 & \textbf{12.18} \\
 & AISHELL-4 & 10.62 & \textbf{10.37} & $-0.25^{\dagger}$ & 10.64 & \textbf{10.38} \\
 & AMI-SDM & 14.01 & \textbf{13.60} & $-0.41^{\dagger}$ & 14.28 & \textbf{13.85} \\
 & NOTSOFAR-1 & 15.74 & \textbf{14.82} & $-0.92^{\dagger}$ & 15.95 & \textbf{15.01} \\
\midrule
\multirow{4}{*}{Predicted} & AliMeeting & 14.05 & \textbf{13.59} & $-0.46^{\dagger}$ & 14.35 & \textbf{13.84} \\
 & AISHELL-4 & 14.00 & \textbf{13.83} & $-0.17^{\dagger}$ & 14.36 & \textbf{14.17} \\
 & AMI-SDM & 16.79 & \textbf{16.07} & $-0.71^{\dagger}$ & 17.60 & \textbf{16.82} \\
 & NOTSOFAR-1 & 21.17 & \textbf{20.34} & $-0.83^{\dagger}$ & 21.84 & \textbf{20.94} \\
\midrule
\multicolumn{2}{@{}l}{Mean RER (\%)} & \multicolumn{3}{c}{3.4} & \multicolumn{2}{c}{3.6} \\
\bottomrule
\end{tabular}%
}
\usebox{\TableBody}
\end{table}

\begin{table*}[!t]
\caption{System Comparison in Our Evaluation on In-Domain and Out-of-Domain Test Sets (\%, Lower Is Better)}
\label{tab:system_comparison}
\centering\footnotesize
\setlength{\tabcolsep}{2pt}
\renewcommand{\arraystretch}{1.05}
\sbox{\TableBody}{%
\begin{tabular*}{\textwidth}{@{}>{\raggedright\arraybackslash}p{78pt}%
@{\extracolsep{0pt plus 1fill}}r@{\extracolsep{0pt plus 1fill}}rr%
@{\extracolsep{0pt plus 3fill}}r@{\extracolsep{0pt plus 1fill}}rr%
@{\extracolsep{0pt plus 3fill}}r@{\extracolsep{0pt plus 1fill}}rr%
@{\extracolsep{0pt plus 3fill}}r@{\extracolsep{0pt plus 1fill}}rr%
@{\extracolsep{0pt plus 5fill}}r@{\extracolsep{0pt plus 1fill}}rr@{}}
\multicolumn{16}{@{}l}{(a) In-Domain}\\[1pt]
\toprule
& \multicolumn{3}{c}{AliMeeting} & \multicolumn{3}{c}{AISHELL-4} & \multicolumn{3}{c}{AMI-SDM} & \multicolumn{3}{c}{NOTSOFAR-1} & \multicolumn{3}{c@{}}{Macro} \\
\cmidrule(lr){2-4}\cmidrule(lr){5-7}\cmidrule(lr){8-10}\cmidrule(lr){11-13}\cmidrule(l){14-16}
Model & cpER & tcpER & DER & cpER & tcpER & DER & cpER & tcpER & DER & cpER & tcpER & DER & cpER & tcpER & DER \\
\midrule
DiCoW(zh/en) & 20.32 & 20.68 & \underline{4.34} & 19.40 & 19.99 & 5.10 & 16.31 & \underline{17.03} & \textbf{8.87} & 20.46 & \underline{21.01} & \underline{5.58} & 19.12 & \underline{19.68} & \textbf{5.97} \\
MOSS-TD & 15.20 & \underline{15.39} & 6.21 & \underline{13.80} & \textbf{14.06} & \textbf{4.27} & 21.84 & 22.77 & 15.02 & 27.32 & 28.24 & 11.61 & 19.54 & 20.12 & 9.28 \\
Qwen3.8-Omni-Flash & 21.74 & 23.04 & 11.33 & 15.54 & 15.84 & 6.08 & 29.32 & 32.25 & 34.08 & 42.60 & 44.22 & 27.01 & 27.30 & 28.84 & 19.62 \\
Proposed (text) & \textbf{13.33} & -- & -- & \textbf{13.67} & -- & -- & \textbf{15.61} & -- & -- & \textbf{19.81} & -- & -- & \textbf{15.60} & -- & -- \\
Proposed (timestamp) & \underline{13.57} & \textbf{13.83} & \textbf{3.91} & 13.84 & \underline{14.19} & \underline{5.02} & \underline{16.07} & \textbf{16.82} & \underline{9.48} & \underline{20.34} & \textbf{20.94} & \textbf{5.52} & \underline{15.96} & \textbf{16.44} & \underline{5.98} \\
\bottomrule
\end{tabular*}%
}
\usebox{\TableBody}\\[6pt]
\sbox{\TableBody}{%
\begin{tabular*}{\textwidth}{@{}>{\raggedright\arraybackslash}p{78pt}%
@{\extracolsep{0pt plus 1fill}}r@{\extracolsep{0pt plus 1fill}}rr%
@{\extracolsep{0pt plus 3fill}}r@{\extracolsep{0pt plus 1fill}}rr%
@{\extracolsep{0pt plus 3fill}}r@{\extracolsep{0pt plus 1fill}}rr%
@{\extracolsep{0pt plus 5fill}}r@{\extracolsep{0pt plus 1fill}}rr@{}}
\multicolumn{13}{@{}l}{(b) Out-of-Domain}\\[1pt]
\toprule
& \multicolumn{3}{c}{MISP-Meeting} & \multicolumn{3}{c}{AISHELL-5} & \multicolumn{3}{c}{DiPCo} & \multicolumn{3}{c@{}}{Macro} \\
\cmidrule(lr){2-4}\cmidrule(lr){5-7}\cmidrule(lr){8-10}\cmidrule(l){11-13}
Model & cpER & tcpER & DER & cpER & tcpER & DER & cpER & tcpER & DER & cpER & tcpER & DER \\
\midrule
DiCoW(zh/en) & 23.33 & 24.08 & 5.37 & 45.68 & 46.17 & 16.20 & 29.24 & 31.25 & \textbf{22.02} & 32.75 & 33.83 & 14.53 \\
MOSS-TD & 17.62 & \underline{17.96} & \underline{4.82} & 30.91 & \underline{31.24} & \textbf{15.35} & \underline{27.60} & \textbf{28.80} & 22.59 & 25.37 & \underline{26.00} & \underline{14.26} \\
Qwen3.8-Omni-Flash & 21.51 & 21.93 & 10.76 & 35.26 & 35.80 & 21.13 & X & X & X & -- & -- & -- \\
Proposed (text) & \textbf{14.04} & -- & -- & \textbf{28.77} & -- & -- & \textbf{26.73} & -- & -- & \textbf{23.18} & -- & -- \\
Proposed (timestamp) & \underline{14.08} & \textbf{14.69} & \textbf{4.74} & \underline{28.84} & \textbf{29.11} & \underline{15.49} & 28.21 & \underline{30.12} & \underline{22.04} & \underline{23.71} & \textbf{24.64} & \textbf{14.09} \\
\bottomrule
\end{tabular*}%
}
\usebox{\TableBody}\\[3pt]
\TableNotes{\textwidth}{Proposed is FireRedAudio + SSC + GSPO. Macro is the equal-weight mean over all datasets within each panel. Bold and underline mark the best and second-best per column. ``--'' marks an unavailable metric, and X marks Qwen3.8-Omni-Flash DiPCo scores left unavailable by one failed API recording. DiCoW and Proposed use S2S refinement priors, whereas MOSS-TD and Qwen3.8-Omni-Flash use no external prior. On MISP-Meeting, target-speaker systems use annotated UEM and MOSS-TD uses the full recording.}
\end{table*}

On clean activity, RSC reduces fixed-target CER from 12.60 to 12.14\%
on AliMeeting and WER from 14.23 to 13.59\% on AMI-SDM.
With target misses, the relative reductions increase to 17.81\% and
14.51\%; with track swaps, they reach 17.84\% and 13.06\%.
Session identity therefore supports target recognition when local activity
loses or exchanges target evidence. False-alarm interventions yield small
increases of 0.02/0.36~pp on AliMeeting/AMI-SDM, indicating that the
benefit centers on missing and exchanged target activity.

The example in Table~\ref{tab:rsc_evidence} illustrates how the same
intervened activity can produce different target transcripts. In a 30-s
AMI window, RSC preserves ``voice recognition'' and the complete reference
phrase, while the activity-only condition yields ``for iso admission''
and an additional word substitution. The supplied activity is identical
in both decodes and contains a 0.43-s target miss. This example shows how session identity changes recognition under the same
input activity.

\subsection{Transfer to FireRedAudio and GSPO Post-Training}
\label{sec:backbone_experiment}
On FireRedAudio, we compare FDDT, SSC, and SSC + GSPO
(Fig.~\ref{fig:fa_progression}). Under the shared 30,000-step SFT recipe, SSC achieves a 2.80\% mean
relative cpER reduction over FDDT across the two priors and four datasets. The SFT-to-GSPO comparison
starts from the same SSC checkpoint and uses the target-error
reward in \eqref{eq:ssc_reward}. GSPO brings a further 3.67\%
mean RER; the combined reduction over FDDT is 6.34\%.
On NOTSOFAR-1, these reductions reach 13.8\% with oracle priors and
9.2\% with predicted priors.

\begin{figure}[!t]
\centering
\includegraphics[width=\columnwidth]{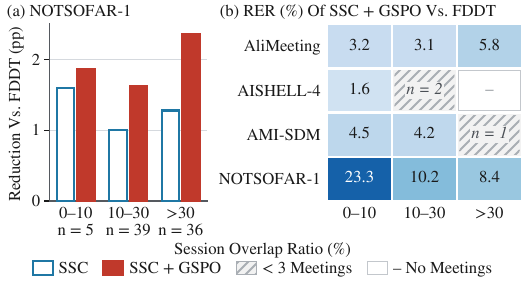}
\caption{FireRedAudio results by full-session overlap ratio with predicted
S2S refinement priors. (a) NOTSOFAR-1 absolute cpER reductions of
SSC and SSC + GSPO from FDDT (pp), with meeting counts $n$.
(b) SSC + GSPO relative
error reduction over FDDT for bins with at least three meetings. Hatched
cells have fewer than three meetings, and dashes mark empty bins.}
\label{fig:session_overlap}
\end{figure}

We group meetings into 0--10\%, 10--30\%, and $>30\%$ overlap.
With predicted priors, SSC + GSPO yields lower cpER than SSC in all
11 nonempty dataset--interval groups. Fig.~\ref{fig:session_overlap}
shows the reductions relative to FDDT, with 5, 39, and 36 NOTSOFAR-1
meetings in the three intervals, respectively. These session-level
comparisons complement the local-concurrency and fully overlapped-target
analyses.

\begin{figure}[!t]
\centering
\includegraphics[width=\columnwidth]{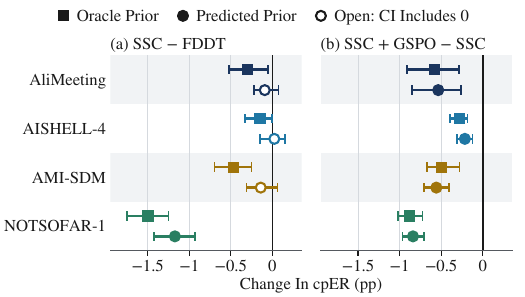}
\caption{Paired meeting-bootstrap changes in corpus-micro cpER for
FireRedAudio: (a) SSC versus FDDT; (b) SSC + GSPO versus SSC.
Squares and circles mark oracle and predicted priors. Open markers
denote CIs spanning zero. Negative values favor the new system.
Error bars show 95\% CIs.}
\label{fig:fa_bootstrap}
\end{figure}

The paired intervals in Fig.~\ref{fig:fa_bootstrap} quantify the
contribution of each stage. Five of eight SSC-versus-FDDT intervals lie entirely below zero.
All eight GSPO-versus-SSC intervals lie below zero, showing consistent
post-training gains across datasets and diarization priors.

\subsection{Sentence Timestamps and Joint Output Evaluation}
\label{sec:timestamp_experiment}
The timestamp variant generates sentence onset/offset times and
text from the same target-conditioned model. Table~\ref{tab:ts_gspo}
compares the SFT checkpoint (SSC) with the checkpoint obtained from it
through GSPO post-training (SSC + GSPO), using the same
target-transcription-error reward. Across the
eight dataset--prior comparisons, mean RER is 3.4\% for cpER and
3.6\% for tcpER. All eight CIs for the cpER differences
(SSC + GSPO minus SSC) lie below zero.
With predicted priors, AMI-SDM tcpER falls from 17.60 to 16.82\%
and NOTSOFAR-1 cpER from 21.17 to 20.34\%.

In the common-protocol system comparison
(Table~\ref{tab:system_comparison}), the timestamp model
incurs an equal-dataset mean cpER cost of 0.43~pp relative to
the text model across seven sets. Its tcpER ranks first among
locally evaluated systems on five sets and second
on AISHELL-4 and DiPCo. Output DER is 3.91\% on AliMeeting and 9.48\% on AMI-SDM.

\subsection{Complete Systems and Generalization to Unseen Domains}
\label{sec:system_experiment}
Table~\ref{tab:system_comparison} compares the complete
FireRedAudio + SSC + GSPO systems with the Chinese--English DiCoW
baseline~\cite{polok2025dicow}, MOSS-TD~\cite{mosi2026moss}, and
Qwen3.8-Omni-Flash~\cite{qwen2026omni38} across the seven test sets.

The proposed text system obtains the lowest cpER among locally
evaluated systems on each of the seven sets. Across the four
in-domain sets, its mean RER is 17.9\% over DiCoW and 17.3\%
over MOSS-TD. Gains over DiCoW are largest on AliMeeting and
AISHELL-4 (34.4\% and 29.5\% RER); gains over MOSS-TD are
largest on AMI-SDM and NOTSOFAR-1 (28.6\% and 27.5\%).
The AISHELL-4 difference from MOSS-TD is smaller, at 0.13~pp.

On the three unseen domains, text-system cpER is
14.04/28.77/26.73\% for MISP-Meeting/AISHELL-5/DiPCo,
compared with DiCoW's 23.33/45.68/29.24\% and MOSS-TD's
17.62/30.91/27.60\%. These gains extend to meeting, in-car, and
dinner-party recordings from new corpora.

\section{Conclusion}
SSC connects local activity structure and session-level identity through
a target condition that controls acoustic feature combinations. TCRE
preserves per-speaker histories and target-centric activity relations.
RSC grounds these relations in acoustic identity derived from session
audio, and CLRM maps the joint condition to diagonal and rank-8
cross-channel modulation for target-speaker recognition. Complete SSC yields 6.23\% mean
RER over STNO-FDDT on Qwen3-ASR-0.6B with 1.46M conditioning
parameters. Transfer to FireRedAudio, followed by
target-error GSPO, produces a complete system with 17.9\% and
17.3\% mean in-domain RER over DiCoW and MOSS-TD,
respectively, and lower cpER than DiCoW and MOSS-TD on three unseen domains.
The timestamp variant retains competitive recognition while
providing sentence-level text and timing in the same output.

\section*{Acknowledgment}
\textbf{Generative AI use.} OpenAI Codex (GPT-6) assisted with translating,
drafting, and polishing author-provided material. Claude Opus 5.5,
accessed through CodeBuddy, assisted with manuscript review and the
preparation of figures and tables.
The authors developed all scientific content, reviewed the manuscript
sentence by sentence, checked and corrected the mathematical derivations
and their presentation, and verified numerical values. They manually
adjusted the AI-drafted visual elements and finalized the manuscript,
figures, and tables, taking full responsibility for the work.

\bibliographystyle{IEEEtran}
\bibliography{references}

@inproceedings{yoshioka2019advances,
  author = {Yoshioka, Takuya and Abramovski, Igor and Aksoylar, Cem and Chen, Zhuo and David, Moshe and Dimitriadis, Dimitrios and Gong, Yifan and Gurvich, Ilya and Huang, Xuedong and Huang, Yan and Hurvitz, Aviv and Jiang, Li and Koubi, Sharon and Krupka, Eyal and Leichter, Ido and Liu, Changliang and Parthasarathy, Partha and Vinnikov, Alon and Wu, Lingfeng and Xiao, Xiong and Xiong, Wayne and Wang, Huaming and Wang, Zhenghao and Zhang, Jun and Zhao, Yong and Zhou, Tianyan},
  title = {Advances in online audio-visual meeting transcription},
  booktitle = {Proc. IEEE Autom. Speech Recognit. Understanding Workshop (ASRU)},
  year = {2019},
  pages = {276--283},
  doi = {10.1109/asru46091.2019.9003827},
}

@inproceedings{vonneumann2024meeting,
  title         = {Meeting Recognition with Continuous Speech Separation and Transcription-Supported Diarization},
  author        = {von Neumann, Thilo and Boeddeker, Christoph and Cord-Landwehr, Tobias and Delcroix, Marc and Haeb-Umbach, Reinhold},
  booktitle = {Proc. IEEE Int. Conf. Acoust., Speech, Signal Process. Workshops (ICASSPW)},
  pages         = {775--779},
  year          = {2024},
  doi           = {10.1109/icasspw62465.2024.10625894},
  url           = {https://doi.org/10.1109/icasspw62465.2024.10625894},
}

@inproceedings{kanda2020joint,
  title = {Joint speaker counting, speech recognition, and speaker identification for overlapped speech of any number of speakers},
  author = {Naoyuki Kanda and Yashesh Gaur and Xiaofei Wang and Zhong Meng and Zhuo Chen and Tianyan Zhou and Takuya Yoshioka},
  booktitle = {Proc. Interspeech},
  year = {2020},
  pages = {36--40},
  doi = {10.21437/Interspeech.2020-1085},
  url = {https://www.isca-archive.org/interspeech_2020/kanda20_interspeech.html},
}

@misc{mosi2026moss,
  title = {{MOSS Transcribe Diarize} technical report},
  author        = {{MOSI.AI} and Yu, Donghua and Lin, Zhengyuan and Chen, Hanfu and Yang, Chen and Zhang, Yiyang and Chen, Jingqi and Chen, Ke and Fan, Liwei and Jiang, Yi and Zhu, Jie and Li, Muchen and Wang, Wenxuan and Wang, Yang and Xu, Zhe and Jiang, Botian and Gong, Yitian and Zhang, Yuqian and Zhang, Wenbo and Wang, Songlin and Wu, Zhiyu and Fei, Zhaoye and Cheng, Qinyuan and Li, Shimin and Qiu, Xipeng},
  year          = {2026},
  eprint        = {2601.01554},
  archivePrefix = {arXiv},
  primaryClass  = {cs.SD},
  note          = {arXiv:2601.01554, version 7},
  url           = {https://arxiv.org/abs/2601.01554v7},
}

@inproceedings{huo2026tagspeech,
  title = {{TagSpeech}: End-to-end multi-speaker {ASR} and diarization with fine-grained temporal grounding},
  author = {Mingyue Huo and Yiwen Shao and Yuheng Zhang},
  booktitle = {Proc. Annu. Meeting Assoc. Comput. Linguistics (ACL)},
  year = {2026},
  pages = {41847--41862},
  doi = {10.18653/v1/2026.acl-long.1938},
  url = {https://aclanthology.org/2026.acl-long.1938/},
}

@inproceedings{polok2024target,
  title     = {Target Speaker {ASR} with {Whisper}},
  author    = {Polok, Alexander and Klement, Dominik and Wiesner, Matthew and Khudanpur, Sanjeev and {\v{C}}ernock{\'y}, Jan and Burget, Luk{\'a}{\v{s}}},
  booktitle = {Proc. IEEE Int. Conf. Acoust., Speech, Signal Process. (ICASSP)},
  year      = {2025},
  pages     = {1--5},
  doi       = {10.1109/ICASSP49660.2025.10887683},
}

@article{polok2025dicow,
  title   = {{DiCoW}: Diarization-conditioned {Whisper} for target speaker automatic speech recognition},
  author  = {Polok, Alexander and Klement, Dominik and Kocour, Martin and Han, Jiangyu and Landini, Federico and Yusuf, Bolaji and Wiesner, Matthew and Khudanpur, Sanjeev and {\v{C}}ernock{\'y}, Jan and Burget, Luk{\'a}{\v{s}}},
  journal = {Comput. Speech Lang.},
  volume  = {95},
  
  year    = {2026},
  doi     = {10.1016/j.csl.2025.101841},
  note = {Art. no. 101841},
}

@InProceedings{radford2023whisper,
  title = 	 {Robust Speech Recognition via Large-Scale Weak Supervision},
  author = {Radford, Alec and Kim, Jong Wook and Xu, Tao and Brockman, Greg and {McLeavey}, Christine and Sutskever, Ilya},
  booktitle = {Proc. Int. Conf. Mach. Learn. (ICML)},
  pages = 	 {28492--28518},
  year = 	 {2023},
  editor = 	 {Krause, Andreas and Brunskill, Emma and Cho, Kyunghyun and Engelhardt, Barbara and Sabato, Sivan and Scarlett, Jonathan},
  volume = 	 {202},
  series = {Proc. Mach. Learn. Res.},
  month = 	 {23--29 Jul},
  publisher =    {PMLR},
  url = 	 {https://proceedings.mlr.press/v202/radford23a.html}
}

@inproceedings{groundingspoken2026,
  title = {Grounding spoken {LLMs} in multi-speaker audio via diarization conditioning},
  author = {Alexander Polok and Samuele Cornell and Sathvik Udupa and Honza {\v{C}}ernock{\'y} and Shinji Watanabe and Luk{\'a}{\v{s}} Burget},
  booktitle = {Proc. Interspeech},
  year = {2026},
  pages = {300--305},
  doi = {10.21437/Interspeech.2026-445},
  url = {https://www.isca-archive.org/interspeech_2026/polok26b_interspeech.html},
}

@inproceedings{polok2026sedicow,
  title     = {{SE-DiCoW}: Self-Enrolled Diarization-Conditioned {Whisper}},
  author    = {Polok, Alexander and Klement, Dominik and Cornell, Samuele and Wiesner, Matthew and {\v{C}}ernock{\'y}, Jan and Khudanpur, Sanjeev and Burget, Luk{\'a}{\v{s}}},
  booktitle = {Proc. IEEE Int. Conf. Acoust., Speech, Signal Process. (ICASSP)},
  year      = {2026},
  url = {https://doi.org/10.1109/ICASSP55912.2026.11461785},
  pages = {16712--16716},
  doi = {10.1109/ICASSP55912.2026.11461785},
}

@inproceedings{hirano2025scsot,
  title = {{SC-SOT}: Conditioning the decoder on diarized speaker information for end-to-end overlapped speech recognition},
  author = {Yuta Hirano and Sakriani Sakti},
  booktitle = {Proc. Interspeech},
  year = {2025},
  pages = {4923--4927},
  doi = {10.21437/Interspeech.2025-1243},
  url = {https://www.isca-archive.org/interspeech_2025/hirano25_interspeech.html},
}

@inproceedings{wang2025selfspeaker,
  title = {Speaker targeting via self-speaker adaptation for multi-talker {ASR}},
  author = {Weiqing Wang and Taejin Park and Ivan Medennikov and Jinhan Wang and Kunal Dhawan and He Huang and Nithin {Rao Koluguri} and Jagadeesh Balam and Boris Ginsburg},
  booktitle = {Proc. Interspeech},
  year = {2025},
  pages = {5498--5502},
  doi = {10.21437/Interspeech.2025-2142},
  url = {https://www.isca-archive.org/interspeech_2025/wang25y_interspeech.html},
}

@article{perez2018film,
  title   = {{FiLM}: Visual Reasoning with a General Conditioning Layer},
  author = {Perez, Ethan and Strub, Florian and de Vries, Harm and Dumoulin, Vincent and Courville, Aaron},
  journal = {Proc. AAAI Conf. Artif. Intell.},
  volume  = {32},
  number  = {1},
  year    = {2018},
  doi     = {10.1609/aaai.v32i1.11671},
  pages = {3942--3951},
}

@misc{zheng2025gspo,
  title         = {Group Sequence Policy Optimization},
  author        = {Zheng, Chujie and Liu, Shixuan and Li, Mingze and Chen, Xiong-Hui and Yu, Bowen and Gao, Chang and Dang, Kai and Liu, Yuqiong and Men, Rui and Yang, An and Zhou, Jingren and Lin, Junyang},
  year          = {2025},
  eprint        = {2507.18071},
  archivePrefix = {arXiv},
  primaryClass  = {cs.LG},
  note          = {arXiv:2507.18071},
  url           = {https://arxiv.org/abs/2507.18071},
}

@inproceedings{yu2017pit,
  author = {Yu, Dong and Kolbæk, Morten and Tan, Zheng-Hua and Jensen, Jesper},
  title = {Permutation invariant training of deep models for speaker-independent multi-talker speech separation},
  booktitle = {Proc. IEEE Int. Conf. Acoust., Speech, Signal Process. (ICASSP)},
  year = {2017},
  pages = {241--245},
  doi = {10.1109/icassp.2017.7952154},
}

@inproceedings{wang2023tfgridnet,
  author = {Wang, Zhong-Qiu and Cornell, Samuele and Choi, Shukjae and Lee, Younglo and Kim, Byeong-Yeol and Watanabe, Shinji},
  title = {{TF-GridNet}: Making time-frequency domain models great again for monaural speaker separation},
  booktitle = {Proc. IEEE Int. Conf. Acoust., Speech, Signal Process. (ICASSP)},
  year = {2023},
  pages = {1--5},
  doi = {10.1109/icassp49357.2023.10094992},
}

@inproceedings{kanda2020sot,
  title         = {Serialized Output Training for End-to-End Overlapped Speech Recognition},
  author        = {Kanda, Naoyuki and Gaur, Yashesh and Wang, Xiaofei and Meng, Zhong and Yoshioka, Takuya},
  booktitle = {Proc. Interspeech},
  pages         = {2797--2801},
  year          = {2020},
  doi           = {10.21437/interspeech.2020-999},
  url           = {https://doi.org/10.21437/interspeech.2020-999},
}

@inproceedings{kanda2022tsot,
  title         = {Streaming Multi-Talker {ASR} with Token-Level Serialized Output Training},
  author        = {Naoyuki Kanda and Jian Wu and Yu Wu and Xiong Xiao and Zhong Meng and Xiaofei Wang and Yashesh Gaur and Zhuo Chen and Jinyu Li and Takuya Yoshioka},
  booktitle = {Proc. Interspeech},
  pages         = {3774--3778},
  year          = {2022},
  doi           = {10.21437/Interspeech.2022-7},
  url           = {https://doi.org/10.21437/Interspeech.2022-7},
}

@inproceedings{subramanian2025segsot,
  title = {Improving practical aspects of end-to-end multi-talker speech recognition for online and offline scenarios},
  author    = {Aswin Shanmugam Subramanian and Amit Das and Naoyuki Kanda and Jinyu Li and Xiaofei Wang and Yifan Gong},
  year      = {2025},
  booktitle = {Proc. Interspeech},
  pages     = {5508--5512},
  doi       = {10.21437/Interspeech.2025-2414},
  issn      = {2958-1796},
}

@inproceedings{shi2024llmsot,
  author = {Shi, Mohan and Jin, Zengrui and Xu, Yaoxun and Xu, Yong and Zhang, Shi-Xiong and Wei, Kun and Shao, Yiwen and Zhang, Chunlei and Yu, Dong},
  title = {Advancing multi-talker {ASR} performance with large language models},
  booktitle = {Proc. IEEE Spoken Lang. Technol. Workshop (SLT)},
  year = {2024},
  pages = {14--21},
  doi = {10.1109/slt61566.2024.10832362},
}

@misc{peng2026vibevoice,
  title         = {{VIBEVOICE-ASR} Technical Report},
  author        = {Zhiliang Peng and Jianwei Yu and Yaoyao Chang and Zilong Wang and Li Dong and Yingbo Hao and Yujie Tu and Chenyu Yang and Wenhui Wang and Songchen Xu and Yutao Sun and Hangbo Bao and Weijiang Xu and Yi Zhu and Zehua Wang and Ting Song and Yan Xia and Zewen Chi and Shaohan Huang and Liang Wang and Chuang Ding and Shuai Wang and Xie Chen and Furu Wei},
  year          = {2026},
  eprint        = {2601.18184},
  archivePrefix = {arXiv},
  primaryClass  = {cs.SD},
  note          = {arXiv:2601.18184},
  url           = {https://arxiv.org/abs/2601.18184},
}

@misc{mu2026diascriber,
  title         = {{DiaScriber}: A Speech {LLM} for Joint Diarization and Transcription in Multi-Speaker Scenarios},
  author        = {Bingshen Mu and Xian Shi and Xiong Wang and Zhifang Guo and Ting He and Xize Cheng and Yu Xi and Jin Xu and Lei Xie},
  year          = {2026},
  eprint        = {2608.22796},
  archivePrefix = {arXiv},
  primaryClass  = {eess.AS},
  note          = {arXiv:2608.22796},
  url           = {https://arxiv.org/abs/2608.22796},
}

@inproceedings{fujita2019eend,
  title = {End-to-end neural speaker diarization with permutation-free objectives},
  author    = {Yusuke Fujita and Naoyuki Kanda and Shota Horiguchi and Kenji Nagamatsu and Shinji Watanabe},
  year      = {2019},
  booktitle = {Proc. Interspeech},
  pages     = {4300--4304},
  doi       = {10.21437/Interspeech.2019-2899},
  issn      = {2958-1796},
}

@inproceedings{eendeda2020,
  title = {End-to-end speaker diarization for an unknown number of speakers with encoder-decoder based attractors},
  author    = {Shota Horiguchi and Yusuke Fujita and Shinji Watanabe and Yawen Xue and Kenji Nagamatsu},
  year      = {2020},
  booktitle = {Proc. Interspeech},
  pages     = {269--273},
  doi       = {10.21437/Interspeech.2020-1022},
  issn      = {2958-1796},
}

@inproceedings{tsvad2020,
  title = {Target-speaker voice activity detection: A novel approach for multi-speaker diarization in a dinner party scenario},
  author    = {Ivan Medennikov and Maxim Korenevsky and Tatiana Prisyach and Yuri Khokhlov and Mariya Korenevskaya and Ivan Sorokin and Tatiana Timofeeva and Anton Mitrofanov and Andrei Andrusenko and Ivan Podluzhny and Aleksandr Laptev and Aleksei Romanenko},
  year      = {2020},
  booktitle = {Proc. Interspeech},
  pages     = {274--278},
  doi       = {10.21437/Interspeech.2020-1602},
  issn      = {2958-1796},
}

@misc{li2026dmasr,
  title         = {{DM-ASR}: Diarization-aware Multi-speaker {ASR} with Large Language Models},
  author        = {Li Li and Ming Cheng and Weixin Zhu and Yannan Wang and Juan Liu and Ming Li},
  year          = {2026},
  eprint        = {2604.22467},
  archivePrefix = {arXiv},
  primaryClass  = {eess.AS},
  note          = {arXiv:2604.22467},
  url           = {https://arxiv.org/abs/2604.22467},
}

@inproceedings{snyder2018xvectors,
  author = {Snyder, David and Garcia-Romero, Daniel and Sell, Gregory and Povey, Daniel and Khudanpur, Sanjeev},
  title = {X-vectors: Robust {DNN} embeddings for speaker recognition},
  booktitle = {Proc. IEEE Int. Conf. Acoust., Speech, Signal Process. (ICASSP)},
  year = {2018},
  pages = {5329--5333},
  doi = {10.1109/icassp.2018.8461375},
}

@inproceedings{desplanques2020ecapa,
  title = {{ECAPA-TDNN}: Emphasized channel attention, propagation and aggregation in {TDNN} based speaker verification},
  author    = {Brecht Desplanques and Jenthe Thienpondt and Kris Demuynck},
  year      = {2020},
  booktitle = {Proc. Interspeech},
  pages     = {3830--3834},
  doi       = {10.21437/Interspeech.2020-2650},
  issn      = {2958-1796},
}

@article{zmolikova2019speakerbeam,
  title         = {{SpeakerBeam}: Speaker Aware Neural Network for Target Speaker Extraction in Speech Mixtures},
  author = {{\v{Z}}mol{\'i}kov{\'a}, Kate{\v{r}}ina and Delcroix, Marc and Kinoshita, Keisuke and Ochiai, Tsubasa and Nakatani, Tomohiro and Burget, Luk{\'a}{\v{s}} and {\v{C}}ernock{\'y}, Jan},
  journal = {IEEE J. Sel. Topics Signal Process.},
  volume        = {13},
  number        = {4},
  pages         = {800--814},
  year          = {2019},
  doi           = {10.1109/jstsp.2019.2922820},
  url           = {https://doi.org/10.1109/jstsp.2019.2922820},
}

@inproceedings{wang2019voicefilter,
  title = {{VoiceFilter}: Targeted voice separation by speaker-conditioned spectrogram masking},
  author    = {Quan Wang and Hannah Muckenhirn and Kevin Wilson and Prashant Sridhar and Zelin Wu and John R. Hershey and Rif A. Saurous and Ron J. Weiss and Ye Jia and Ignacio Lopez Moreno},
  year      = {2019},
  booktitle = {Proc. Interspeech},
  pages     = {2728--2732},
  doi       = {10.21437/Interspeech.2019-1101},
  issn      = {2958-1796},
}

@inproceedings{kanda2019simultaneous,
  author = {Kanda, Naoyuki and Horiguchi, Shota and Fujita, Yusuke and Xue, Yawen and Nagamatsu, Kenji and Watanabe, Shinji},
  title = {Simultaneous speech recognition and speaker diarization for monaural dialogue recordings with target-speaker acoustic models},
  booktitle = {Proc. IEEE Autom. Speech Recognit. Understanding Workshop (ASRU)},
  year = {2019},
  pages = {31--38},
  doi = {10.1109/asru46091.2019.9004009},
}

@article{sqwhisper2024,
  title         = {{SQ-Whisper}: Speaker-Querying Based {Whisper} Model for Target-Speaker {ASR}},
  author        = {Guo, Pengcheng and Chang, Xuankai and Lv, Hang and Watanabe, Shinji and Xie, Lei},
  journal = {IEEE Trans. Audio, Speech, Lang. Process.},
  volume        = {33},
  pages         = {175--185},
  year          = {2025},
  doi           = {10.1109/taslp.2024.3513835},
  url           = {https://doi.org/10.1109/taslp.2024.3513835},
}

@inproceedings{zaheer2017deepsets,
  title         = {{Deep Sets}},
  author = {Zaheer, Manzil and Kottur, Satwik and Ravanbakhsh, Siamak and Poczos, Barnabas and Salakhutdinov, Ruslan and Smola, Alexander J.},
  booktitle     = {Advances in Neural Information Processing Systems},
  volume        = {30},
  year          = {2017},
  url = {https://proceedings.neurips.cc/paper/2017/hash/f22e4747da1aa27e363d86d40ff442fe-Abstract.html},
}

@inproceedings{santoro2017relations,
  title         = {A simple neural network module for relational reasoning},
  author        = {Santoro, Adam and Raposo, David and Barrett, David G and Malinowski, Mateusz and Pascanu, Razvan and Battaglia, Peter and Lillicrap, Timothy},
  booktitle     = {Advances in Neural Information Processing Systems},
  volume        = {30},
  year          = {2017},
  url           = {https://proceedings.neurips.cc/paper_files/paper/2017/file/e6acf4b0f69f6f6e60e9a815938aa1ff-Paper.pdf},
}

@inproceedings{okabe2018attentive,
  title = {Attentive statistics pooling for deep speaker embedding},
  author    = {Koji Okabe and Takafumi Koshinaka and Koichi Shinoda},
  year      = {2018},
  booktitle = {Proc. Interspeech},
  pages     = {2252--2256},
  doi       = {10.21437/Interspeech.2018-993},
  issn      = {2958-1796},
}

@inproceedings{ha2017hypernetworks,
  title         = {{HyperNetworks}},
  author        = {David Ha and Andrew Dai and Quoc V. Le},
  booktitle     = {Proc. Int. Conf. on Learning Representations (ICLR)},
  year          = {2017},
  eprint        = {1609.09106},
  archivePrefix = {arXiv},
  primaryClass  = {cs.LG},
  url           = {https://arxiv.org/abs/1609.09106},
}

@inproceedings{zhao2016lrpd,
  title         = {Low-rank plus diagonal adaptation for deep neural networks},
  author        = {Zhao, Yong and Li, Jinyu and Gong, Yifan},
  booktitle = {Proc. IEEE Int. Conf. Acoust., Speech, Signal Process. (ICASSP)},
  pages         = {5005--5009},
  year          = {2016},
  doi           = {10.1109/icassp.2016.7472630},
  url           = {https://doi.org/10.1109/icassp.2016.7472630},
}

@InProceedings{houlsby2019adapters,
  title = 	 {Parameter-Efficient Transfer Learning for {NLP}},
  author =       {Houlsby, Neil and Giurgiu, Andrei and Jastrzebski, Stanislaw and Morrone, Bruna and De Laroussilhe, Quentin and Gesmundo, Andrea and Attariyan, Mona and Gelly, Sylvain},
  booktitle = {Proc. Int. Conf. Mach. Learn. (ICML)},
  pages = 	 {2790--2799},
  year = 	 {2019},
  editor = 	 {Chaudhuri, Kamalika and Salakhutdinov, Ruslan},
  volume = 	 {97},
  series = {Proc. Mach. Learn. Res.},
  month = 	 {09--15 Jun},
  publisher =    {PMLR},
  url = 	 {https://proceedings.mlr.press/v97/houlsby19a.html}
}

@inproceedings{hu2022lora,
  title         = {{LoRA}: Low-Rank Adaptation of Large Language Models},
  author        = {Edward J. Hu and Yelong Shen and Phillip Wallis and Zeyuan Allen-Zhu and Yuanzhi Li and Shean Wang and Lu Wang and Weizhu Chen},
  booktitle     = {Proc. Int. Conf. on Learning Representations (ICLR)},
  year          = {2022},
  eprint        = {2106.09685},
  archivePrefix = {arXiv},
  primaryClass  = {cs.CL},
  url           = {https://arxiv.org/abs/2106.09685},
}

@inproceedings{yu2022m2met,
      title = {{M2MeT}: The {ICASSP} 2022 multi-channel multi-party meeting transcription challenge}, 
      author={Fan Yu and Shiliang Zhang and Yihui Fu and Lei Xie and Siqi Zheng and Zhihao Du and Weilong Huang and Pengcheng Guo and Zhijie Yan and Bin Ma and Xin Xu and Hui Bu},
      year={2022},
      
      
      
      url = {https://doi.org/10.1109/ICASSP43922.2022.9746465},
  booktitle = {Proc. IEEE Int. Conf. Acoust., Speech, Signal Process. (ICASSP)},
  pages = {6167--6171},
  doi = {10.1109/ICASSP43922.2022.9746465},
}

@inproceedings{fu2021aishell4,
  title = {{AISHELL-4}: An open source dataset for speech enhancement, separation, recognition and speaker diarization in conference scenario},
  author    = {Yihui Fu and Luyao Cheng and Shubo Lv and Yukai Jv and Yuxiang Kong and Zhuo Chen and Yanxin Hu and Lei Xie and Jian Wu and Hui Bu and Xin Xu and Jun Du and Jingdong Chen},
  year      = {2021},
  booktitle = {Proc. Interspeech},
  pages     = {3665--3669},
  doi       = {10.21437/Interspeech.2021-1397},
  issn      = {2958-1796},
}

@incollection{carletta2006ami, title = {The {AMI} meeting corpus: A pre-announcement}, ISBN={9783540325505}, ISSN={1611-3349}, url = {https://doi.org/10.1007/11677482_3}, DOI={10.1007/11677482_3}, booktitle={Machine Learning for Multimodal Interaction}, publisher={Springer Berlin Heidelberg}, author={Carletta, Jean and Ashby, Simone and Bourban, Sebastien and Flynn, Mike and Guillemot, Mael and Hain, Thomas and Kadlec, Jaroslav and Karaiskos, Vasilis and Kraaij, Wessel and Kronenthal, Melissa and Lathoud, Guillaume and Lincoln, Mike and Lisowska, Agnes and McCowan, Iain and Post, Wilfried and Reidsma, Dennis and Wellner, Pierre}, year={2006}, pages={28--39} }

@inproceedings{vinnikov2024notsofar,
  title = {{NOTSOFAR-1} challenge: New datasets, baseline, and tasks for distant meeting transcription},
  author    = {Alon Vinnikov and Amir Ivry and Aviv Hurvitz and Igor Abramovski and Sharon Koubi and Ilya Gurvich and Shai Peer and Xiong Xiao and Benjamin Martinez Elizalde and Naoyuki Kanda and Xiaofei Wang and Shalev Shaer and Stav Yagev and Yossi Asher and Sunit Sivasankaran and Yifan Gong and Min Tang and Huaming Wang and Eyal Krupka},
  year      = {2024},
  booktitle = {Proc. Interspeech},
  pages     = {5003--5007},
  doi       = {10.21437/Interspeech.2024-1788},
  issn      = {2958-1796},
}

@inproceedings{gao2025misp,
  title = {The multimodal information based speech processing ({MISP}) 2025 challenge: Audio-visual diarization and recognition},
  author    = {Ming Gao and Shilong Wu and Hang Chen and Jun Du and Chin-Hui Lee and Shinji Watanabe and Jingdong Chen and Sabato Marco Siniscalchi and Odette Scharenborg},
  year      = {2025},
  booktitle = {Proc. Interspeech},
  pages     = {1888--1892},
  doi       = {10.21437/Interspeech.2025-2017},
  issn      = {2958-1796},
}

@inproceedings{dai2025aishell5,
  title = {{AISHELL-5}: The first open-source in-car multi-channel multi-speaker speech dataset for automatic speech diarization and recognition},
  author    = {Yuhang Dai and He Wang and Xingchen Li and Zihan Zhang and Shuiyuan Wang and Lei Xie and Xin Xu and Hongxiao Guo and Shaoji Zhang and Hui Bu and Wei Chen},
  year      = {2025},
  booktitle = {Proc. Interspeech},
  pages     = {5493--5497},
  doi       = {10.21437/Interspeech.2025-1886},
  issn      = {2958-1796},
}

@inproceedings{vansegbroeck2019dipco,
      title = {{DiPCo}---Dinner party corpus}, 
      author = {{Van Segbroeck}, Maarten and Zaid, Ahmed and Kutsenko, Ksenia and Huerta, Cirenia and Nguyen, Tinh and Luo, Xuewen and Hoffmeister, Bj{\"o}rn and Trmal, Jan and Omologo, Maurizio and Maas, Roland},
      year = {2020},
      
      
      
      url = {https://www.isca-archive.org/interspeech_2020/segbroeck20_interspeech.html},
  booktitle = {Proc. Interspeech},
  pages = {434--436},
  doi = {10.21437/Interspeech.2020-2800},
}

@inproceedings{han2025diarizen,
  title = {Leveraging self-supervised learning for speaker diarization},
  author = {Jiangyu Han and Federico Landini and Johan Rohdin and Anna Silnova and Mireia Diez and Luk{\'a}{\v{s}} Burget},
  booktitle = {Proc. IEEE Int. Conf. Acoust., Speech, Signal Process. (ICASSP)},
  year = {2025},
  pages = {1--5},
  doi = {10.1109/ICASSP49660.2025.10889475},
  url = {https://doi.org/10.1109/ICASSP49660.2025.10889475},
}

@misc{wang2026s2srefinement,
  author = {Ruoyu Wang and Jun Du and Weixin Zhu and Yannan Wang},
  title = {Sequence-to-Sequence Refinement for {EEND-VC} Diarization with a Shared Encoder},
  year = {2026},
  eprint = {2611.02709},
  archivePrefix = {arXiv},
  primaryClass = {eess.AS},
  note = {arXiv:2611.02709},
  url = {https://arxiv.org/abs/2611.02709},
}

@misc{qwen2026asr,
      title = {{Qwen3-ASR} technical report}, 
      author={Xian Shi and Xiong Wang and Zhifang Guo and Yongqi Wang and Pei Zhang and Xinyu Zhang and Zishan Guo and Hongkun Hao and Yu Xi and Baosong Yang and Jin Xu and Jingren Zhou and Junyang Lin},
      year={2026},
      eprint={2601.21337},
      archivePrefix={arXiv},
      primaryClass={cs.CL},
      url={https://arxiv.org/abs/2601.21337},
  note = {arXiv:2601.21337},
}

@misc{shen2026fireredaudio,
      title = {{FireRedAudio}: A general-purpose audio language model with decoupled continuous representations for understanding and generation}, 
      author={Feiyu Shen and Fenglong Xie and Junjie Li and Kun Xie and Lei Xie and Xu Tang and Xuelong Geng and Yan Jia and Yao Hu and Yichen Han and Yichen Wu and Ziqi Dai and Junjie Chen and Kai Huang and Manzhen Wei and Yixuan Li},
      year={2026},
      eprint={2608.24168},
      archivePrefix={arXiv},
      primaryClass={cs.CL},
      url={https://arxiv.org/abs/2608.24168},
  note = {arXiv:2608.24168},
}

@inproceedings{neumann2023meeteval,
  title = {{MeetEval}: A toolkit for computation of word error rates for meeting transcription systems},
  author    = {Thilo {von Neumann} and Christoph Boeddeker and Marc Delcroix and Reinhold Haeb-Umbach},
  year      = {2023},
  booktitle = {Proc. Int. Workshop Speech Process. Everyday Environ. (CHiME)},
  pages     = {27--32},
  doi       = {10.21437/CHiME.2023-6},
}

@InProceedings {bisani2004bootstrap,
  author= {Bisani, M. and Ney, H.},
  title = {Bootstrap estimates for confidence intervals in {ASR} performance evaluation},
  booktitle = {Proc. IEEE Int. Conf. Acoust., Speech, Signal Process. (ICASSP)},
  year= 2004,
  pages = {409--412},
  address= {Montreal, Canada},
  month= may,
  volume= {1},
  booktitlelink= {http://www.icassp2004.org/},
  doi = {10.1109/ICASSP.2004.1326009},
  url = {https://doi.org/10.1109/ICASSP.2004.1326009},
}

@misc{qwen2026omni38,
  author = {{Qwen Team}},
  title = {{Qwen3.8-Omni}: Towards Native Omni-Modal Agents},
  year = {2026},
  note = {arXiv:2609.25611},
  url = {https://arxiv.org/abs/2609.25611},
}

\end{document}